\documentclass[aps,prd,superscriptaddress,twocolumn,floatfix,10pt,longbibliography]{revtex4-2}
\RequirePackage{snapshot}
\usepackage[dvipsnames]{xcolor}
\definecolor{linkcolor}{rgb}{0.0,0.3,0.5}
\usepackage[unicode, colorlinks=true, linkcolor=linkcolor, citecolor=linkcolor, filecolor=linkcolor, urlcolor=linkcolor, linktocpage, breaklinks]{hyperref}
\usepackage[all]{hypcap}

\usepackage{comment}
\usepackage{amssymb,amsmath,amsthm,latexsym,mathrsfs,verbatim,
            mathtools,needspace,enumitem,etoolbox,graphicx,physics,
            microtype,afterpage,xspace,tabularx,lmodern,multirow,slashed,svg}
\usepackage{orcidlink}
\usepackage{textcomp, gensymb}
\usepackage{flushend} 
\usepackage[T1]{fontenc}
\usepackage[utf8]{inputenc}
\hypersetup{colorlinks=true,citecolor=romared,linkcolor=romared,urlcolor=romared}
\usepackage{environ}
\usepackage{csvsimple,etoolbox}
\usepackage{tikz}
\usepackage{tensor}
\usepackage{placeins} %
\usepackage{longtable}

\definecolor{romared}{RGB}{142,0,28}

\newcommand{\be}{\begin{equation}}
\newcommand{\ee}{\end{equation}}

\def\be{\begin{equation}}
\def\ee{\end{equation}}
\newcommand{\beq}{\begin{eqnarray}}
\newcommand{\eeq}{\end{eqnarray}}

\usepackage{makecell}
\usepackage{soul}

\newcolumntype{Y}{>{\centering\arraybackslash}X}

\usepackage[normalem]{ulem}
\usepackage{adjustbox}
\usepackage{etoolbox}

\begin{document}

\title{Twist and higher modes of a complex scalar field at the threshold of collapse}

\author{Krinio Marouda\orcidlink{0000-0003-1030-8853}} 
\affiliation{CENTRA, Departamento de F\'{\i}sica, Instituto Superior T\'ecnico -- IST, Universidade de Lisboa -- UL, Avenida Rovisco Pais 1, 1049-001 Lisboa, Portugal}

\author{Daniela Cors\orcidlink{0000-0002-0520-2600}} 
    \affiliation{Department of Applied Mathematics and Theoretical Physics, Centre for Mathematical Sciences, University of Cambridge, Wilberforce Road, Cambridge CB3 0WA, United Kingdom}
    
\author{Hannes R. R\"uter\orcidlink{0000-0002-3442-5360}} 
\affiliation{CENTRA, Departamento de F\'{\i}sica, Instituto Superior T\'ecnico -- IST, Universidade de Lisboa -- UL, Avenida Rovisco Pais 1, 1049-001 Lisboa, Portugal}

\author{Alex Va\~no-Vi\~nuales\,\orcidlink{0000-0002-8589-006X}}
\affiliation{CENTRA, Departamento de F\'{\i}sica, Instituto Superior T\'ecnico -- IST, Universidade de Lisboa -- UL, Avenida Rovisco Pais 1, 1049-001 Lisboa, Portugal}
\affiliation{Departament de Física, Universitat de les Illes Balears, IAC3, Carretera Valldemossa km 7.5, E-07122 Palma, Spain}

\author{David Hilditch\orcidlink{0000-0001-9960-5293}} 
\affiliation{CENTRA, Departamento de F\'{\i}sica, Instituto Superior T\'ecnico -- IST, Universidade de Lisboa -- UL, Avenida Rovisco Pais 1, 1049-001 Lisboa, Portugal}

\date{\today}

\begin{abstract}
We investigate the threshold of collapse of a massless complex scalar field in axisymmetric spacetimes under the ansatz of Choptuik \textit{et al.}~\cite{choptuik2004critical}, in which a symmetry depending on the azimuthal parameter~$m$ is imposed on the scalar field. This allows for both non-vanishing twist and angular momentum. We extend earlier work to include higher angular modes. Using the pseudospectral code \textsc{bamps} with a new adapted symmetry reduction method, which we call~$m$-cartoon, and a generalized twist-compatible apparent horizon finder, we evolve near-critical initial data to the verge of black hole formation for the lowest nontrivial modes, $m=1$ and~$m=2$. For~$m=1$ we recover discrete self-similarity with echoing period~$\Delta\simeq0.42$ and power-law scaling with exponent~$\gamma\simeq0.11$, consistent with earlier work. For~$m=2$ we find that universality is maintained within this nonzero fixed-$m$ symmetry class but with smaller period and critical exponents, $\Delta\simeq0.09$ and $\gamma\simeq0.035$, establishing an explicit dependence of the critical solution on the angular mode. Analysis of the relation between the angular momentum and the mass of apparent horizons at the instant of formation, $J_{\mathrm{AH}}{-}M_{\mathrm{AH}}$, shows that the effect of angular momentum is minimal at the threshold, with~$\chi_{\mathrm{AH}}=J_{\mathrm{AH}}/M_{\mathrm{AH}}^2\to0$, and, therefore, excludes extremal black holes for the families under consideration. This conclusion is further validated by the negative subdominant Lyapunov exponent~$\lambda_1$, related to rotations, implying the angular momentum scales down faster than the mass as the threshold is approached. In the presence of angular momentum, $m>0$, we observe a single center of collapse, with quantitative evidence from curvature invariants indicating no competition between the scalar and vacuum thresholds, in contrast to what we witnessed in the absence of twist~\cite{MaroudaCors2024}. Our results demonstrate that while universality and discrete self-similarity hold within each~$m$-sector, the critical universal values vary with~$m$, and neither extremality nor bifurcation occur in the complex scalar field model within the families considered here.
\end{abstract}
 
\maketitle

\section{Introduction}
\label{sec:introduction}

In the three decades following Choptuik's breakthrough paper~\cite{choptuik1993universality}, the threshold of gravitational collapse in general relativity (GR) has been found to exhibit fascinating phenomena in a variety of different physical scenarios. The Choptuik solution itself was found empirically to be {\it the} solution that lies between collapse and dispersion of a smooth massless scalar field minimally coupled to GR in spherical symmetry. In this context spacetimes that are sufficiently weak to disperse are called subcritical, whereas those with initial data strong enough to collapse to a black hole are called supercritical. Solutions that are close enough in solution space to the threshold exhibit universal power-law scaling against the phase space distance from the critical solution. On the supercritical side such scaling is found in the mass of the black holes formed. On the subcritical side curvature invariants are found to scale~\cite{Garfinkle:1998va}. Such spacetimes are said to lie in the critical regime. The Choptuik solution itself exhibits discrete self-similarity (DSS). A DSS spacetime is one that admits coordinates~$T, x^i$ in which the metric can be expressed as
\begin{align}
\label{eq:DSS1}
    g_{ab} = e^{-2T}\,\tilde{g}_{ab}(T, x^i)\,,
\end{align}
with~$\tilde{g}_{ab}(T, x^i)$ a conformal metric that is periodic in~$T$ 
with period~$\Delta$,
\begin{align}
\label{eq:DSS2}
    \tilde{g}_{ab}(T, x^i)=\tilde{g}_{ab}(T+\Delta, x^i)\,,
\end{align}
where the variable $T$ can be any time coordinate adapted to the DSS symmetry. A specific version of this time coordinate is usually referred to as ``slow-time'', defined as
\begin{align}
\label{eq:slowT}
    T=-\ln\left(\tau^*-\tau\right)\,,
\end{align}
with $\tau^*$ being the proper time of accumulation of DSS features. 

These results have been found independent of particular families of initial data, gauge, formulation of GR, discretization methods and so forth. As a result, nowadays it serves as a testing ground for many numerical relativity codes that involve scalar fields. These properties manifest within a large set of matter models in spherical symmetry, and are usually referred to as \textit{`type II critical phenomena in gravitational collapse'}, due to their resemblance to phase transitions in statistical physics~\cite{Gundlach:2025yje, gundlach2007critical, KoikeHara99}.

Studies have indicated a more complicated story as we depart from spherical symmetry, including a potential loss of universality, at least for the case of vanishing angular momentum. Examples of these departures in twist-free axisymmetry concern real scalar fields~\cite{choptuik2003critical, baumgarte2018aspherical}, complex scalar fields~\cite{MaroudaCors2024}, electromagnetic waves~\cite{Baumgarte:2019fai,Mendoza:2021nwq,Reid:2023jmr} and gravitational waves~\cite{SuarezFernandez:2022hyx, Ledvinka:2021rve, baumgarte2023critical}. A very different numerical setup for axisymmetric scalar field collapse in single-null coordinates, whose results do not completely fit with the remainder, is given in~\cite{Gundlach:2024eds}. It is tempting to interpret this more complicated phenomenology as a consequence of the competition between the threshold solutions of the matter and dynamical gravitational waves, as the latter inevitably arise in axisymmetry. Thus, as a natural continuation of earlier work, we study here the case of a complex scalar field with angular momentum, thus extending our earlier work~\cite{MaroudaCors2024} to include the most general setup available in axisymmetry. This is achieved by working under an ansatz introduced in~\cite{choptuik2004critical}, namely
\begin{align}
\label{eq:ansatz}
    \psi\left(\rho,\phi,z,t\right) = e^{i m \phi} \psi_m(\rho,z,t)\,.
\end{align}
in standard cylindrical polar coordinates. As a matter of fact, introducing angular momentum to the spacetime with scalar field matter content yields neither a spherically symmetric nor an obvious vacuum limit case that could serve as a basis for comparison of the threshold solutions. More precisely, no spherical limit exists in this case, in the sense that even perturbative solutions within our symmetry class are not spherical. Likewise, in vacuum axisymmetry, black holes with angular momentum cannot be formed from regular initial data (see App.~\ref{app:no_competitor} for details). In contrast to the~$m=0$ case~\cite{MaroudaCors2024}, when adding net angular momentum in this manner and fixing the azimuthal winding number~$m=1$, it was found in~\cite{choptuik2004critical} that the threshold of collapse is again rendered universal, for numerous families of~$l=|m|=1$ initial data. This part of the solution space is a particularly interesting, since it is not spherical yet displays critical behavior familiar from that setting. Studying it in detail may therefore strengthen our understanding of critical phenomena and shed light on the loss of universality at the axisymmetric threshold of collapse. Presently we look at different families of initial data, extending to the higher angular mode~$m=2$.

The axisymmetric formation of Kerr black holes is moreover of great interest on its own since it corresponds to the creation of the most generic, simple and unique stationary astrophysical compact objects we know. Recently, there has been a new perspective added to the picture of critical phenomena by Kehle and Unger~\cite{kehle2024extremal, kehle2024gravitationalcollapseextremalblack}, who proved not only that extremal black holes can exist in collapse scenarios, but that they occur in the limit of marginal formation. They studied in particular the spherical charged Einstein-Maxwell-Vlasov system, and it was proven that for a carefully chosen family of initial data, an extremal Reissner-Nordström solution lies at the threshold of collapse. They furthermore showed that extremal Reissner-Nordström black hole formation also occurs in simpler models, such as bouncing charged null dust and thin charged shells, reinforcing that the picture of ``extremal critical collapse'' is not unique to one system. Charge in black hole spacetimes is regularly used as a proxy for the effect of angular momentum, so these results lead naturally to conjectures on the latter. Two striking conjectures they make concern the existence of an extremal Reissner-Nordsröm black hole at the threshold of collapse in the spherical Einstein-Maxwell charged scalar field model, and the existence of an extremal Kerr black hole at the threshold in vacuum in 3+1-dimensions. Since black holes formed in the collapse of complex scalar fields with angular momentum in axisymmetry are generically expected to settle to Kerr, one may suspect that this conjecture could extend to the occurrence of extremal Kerr black holes at the threshold within the Einstein-Klein-Gordon system with angular momentum as we study in this paper. Simulated experiments could then provide evidence for the validity of such conjectures. In particular, in case of a tendency towards extremality as the black hole threshold is approached by our simulations would serve as numerical support of this extension to their conjecture. As already shown in~\cite{choptuik2004critical} for~$m=1$ and confirmed in what follows for~$m=2$, however, no such trend is observed within our data. The possibility of extremal Kerr at the threshold of the axisymmetric Einstein-Klein-Gordon system therefore remains open, but may require very different families of initial data or perhaps the inclusion of a non-zero mass term.

From the numerical relativity point of view, the scheme we present below renders possible the examination of the threshold of Kerr black hole formation for the first time in a pseudospectral code. In the case of~\cite{choptuik2004critical}, which we will reproduce and extend, this was achieved by plugging in the ansatz~\eqref{eq:ansatz} into the equations and solving for the~$\phi$ independent part of the ansatz, $\psi_m(\rho,z,t)$.
The authors of~\cite{choptuik2004critical} evolved the resulting system of equations with a finite-differencing code. Within \textsc{bamps}, the code we are using, it was instead necessary to generalize the cartoon method for symmetry reduction and to adapt our horizon finder~\cite{ahloc3d} for the post-process classification of the supercritical side, in order to implement the same system efficiently in axisymmetry.

The paper is structured as follows. An overview of the equations of motion, the initial-data solver and code setup is given in sections~\ref{sec:massless_complex_scalar_field} and~\ref{subsection:Formulation}-\ref{subsec:hprefinement}. The aforementioned improvements to symmetry reduction and our horizon finder are explained in sections~\ref{subsec:cartoom} and~\ref{subsec:phase_space_search}. Our numerical results are discussed in section~\ref{sec:numerical_results}. We conclude in section~\ref{sec:conclusions}. In appendix~\ref{app:no_competitor} we give a concise summary of the relationship between angular momentum in vacuum collapse spacetimes in axisymmetry.
 
\section{Massless complex scalar field with angular momentum}
\label{sec:massless_complex_scalar_field}

The matter model under consideration is the same as in our previous work~\cite{MaroudaCors2024}, whereas the different symmetry class, see Eq.~\eqref{eq:ansatz}, alters the solution space. Assuming geometric units $G=c=1$, the action is given by
\begin{align}
    S=\int \mathrm{d}^4 x \sqrt{-g} \left(\frac{R}{16\pi}-\frac{1}{2}\nabla_a \psi \nabla^a \psi^\dagger \right)\, ,
\end{align}
where~$R$ is the Ricci scalar,~$g$ is the determinant of the metric,~$\psi=\psi^{\textrm{Re}}+i \psi^{\textrm{Im}}$ is the complex scalar field and~$\psi^\dagger$ its complex conjugate. Latin indices starting from~$a$ denote spacetime components, while indices starting from~$i$ refer to the spatial components in standard 3+1 coordinates. The equations of motion corresponding to this action are Einstein's field equations coupled to the massless Klein-Gordon equation, 
\begin{align}
    G_{ab} &=R_{ab}-\frac{1}{2}g_{ab}R =8\pi T_{ab}\,,\label{eqn:EEs}\\
    \square \psi &= g^{ab} \nabla_a \nabla_b \psi =0\,.
\end{align}
where the stress-energy tensor is the following
\begin{align}
    \label{eq:Tab}
    T_{ab}=\nabla_{\left(a\right.} \psi  \nabla_{\left.b\right)}\psi^\dagger -\frac{1}{2}g_{ab} \nabla^c \psi \nabla_c \psi^\dagger\,.
\end{align}
In this study, we impose the ansatz~\eqref{eq:ansatz} of a single angular mode to the scalar field in the initial data and throughout the evolution. This ansatz does not imply a~$\phi$-dependence on the metric, which maintains the axisymmetry of the spacetime. Simultaneously, the asymmetry ($\phi$-dependence) of the scalar field~(Eq.~\eqref{eq:ansatz}) is chosen in such a way that it renders certain metric components non-zero. Those cross terms are responsible for the presence of net Komar angular momentum on the slice, which, in the absence of a trapped surface (see Appendix~\ref{app:no_competitor}) corresponds to
\begin{align}
\label{eq:J_spacetime}
    J=-\frac{1}{8\pi}\int_{\Sigma_t} T_{ab}\xi^a n^b \sqrt{\det\gamma} \, \mathrm{d}^3x\,,
\end{align}
where~$\Sigma_t$ is the spacelike slice of constant~$t$ in our coordinates, $\det \gamma$ is the determinant of the induced metric on it, $n^a$ is the timelike hypersurface normal and~$\xi^a$ is the Killing vector associated with the azimuthal symmetry of the spacetime,
\begin{align}
\label{eq:Killing}
    \xi^a=(\partial_\phi)^a=x(\partial_y)^a-y(\partial_x)^a\,,
\end{align}
where for later convenience we recall the relationship between the Killing vector and standard Cartesian partial derivatives.

The twist 4-vector is defined as 
\begin{align}
    \omega_a = \epsilon_{abcd}\,\xi^b \nabla^c \xi^d\,,
\end{align}
where $\epsilon_{abcd}$ is the 4D Levi-Civita antisymetric tensor and $\nabla^c$ is the covariant derivative compatible with the metric $g_{ab}$~\cite{rinne2013axisymmetric}. In the reduced 3-manifold after symmetry reduction, components of the twist appear in the evolution equations and carry information about the ``rotation'' of the Killing field. In fact, twist measures how the Killing vector of axisymmetry fails to be hypersurface orthogonal, so it ``twists'' around itself and it is physically associated with rotational/axial degrees of freedom in the spacetime. Note that a spacetime that has angular momentum will have a non-zero twist, but a non-zero twist does not guarantee a non-zero angular momentum. For more details refer to Appendix~\ref{app:no_competitor}, in particular consider Eq.~\eqref{eq:Twist_J} for the former and Eq.~\eqref{eq:J_M+BH} without matter or trapped surfaces for the latter case.

\section{Formulation and numerical methods}
\label{sec:formulation_and_numerical_methods}

In order to implement this system numerically in axisymmetry we have developed a generalized symmetry reduction method in \textsc{bamps}, adapted to the symmetries of the problem. The \textsc{bamps} code is a pseudospectral code for the time development of first-order symmetric hyperbolic systems. The name chosen for the new symmetry reduction method is \textit{m-cartoon method}. It is a generalization of the \textit{cartoon method}, simply adapted to spacetimes where there is a net angular momentum in the complex scalar field content. In this section we give a brief summary of the continuum equations solved, our grid setup, initial data solver, $hp$-mesh-refinement strategy and our improved constraint damping parameters. Full details regarding these aspects of the code can be found in the technical references~\cite{Bruegmann:2011zj,hilditch2016pseudospectral,renkhoff2023adaptive,cors2023formulation}. Following these, a detailed explanation of the novel \textit{m-cartoon method} is provided. We end the section with an overview of the adjustments we made to our apparent horizon finder to include twist.

\subsection{Formulation of General Relativity}
\label{subsection:Formulation}

\subsubsection{Geometry}
\label{subsubsection:Geometry}

The formulation of GR used in the present work is identical to that in our
previous work~\cite{MaroudaCors2024} where generalized harmonic gauge (GHG) is imposed~\cite{Pretorius:2004jg, Lindblom:2005qh}, and then, first derivatives of metric components are chosen as evolved quantities in such a way that a first-order symmetric hyperbolic formulation of GR is ensured.
The system is enhanced by a damping scheme for the harmonic constraints,
\begin{align}
C_a=\Gamma_a+H_a=0\,,
\end{align} inspired by~\cite{Gundlach:2005eh}. Here,~$\Gamma^a=g^{bc}\Gamma^a{}_{bc}$, is the contracted Christoffel symbol with the inverse metric The specific choice of gauge source functions~$H_a$ for this work is the DSS-compatible gauge choice introduced
in~\cite{cors2023formulation} and used in~\cite{MaroudaCors2024} as it respects necessary conditions to adapt to the symmetry during the DSS phase of the evolutions near the threshold of collapse, 
\begin{align}
    H_a \left(T+\Delta,x^i\right)=H_a \left(T,x^i\right)\,,
\end{align}
where~$\Delta$ is the DSS-period of the features of collapse at the threshold of black hole formation. For the parameters of that gauge, we select the exact same values $\eta_L=4$ and $\eta_S=6$, see~\cite{cors2023formulation,MaroudaCors2024} for the relevant expression.

The evolved variables are~$g_{ab}$, $\Pi_{ab}$ and~$\Phi_{iab}$ and the evolution equations as stated in~\cite{MaroudaCors2024}. Here, $\Pi_{ab}=-n^c\partial_c g_{ab}$ is the time reduction variable and~$\Phi_{iab}=\partial_i g_{ab}$ is the spatial reduction variable. The reduction constraint related to the spatial derivatives is~$C_{iab}:=\partial_i g_{ab}-\Phi_{iab}$, which is damped according to the general prescription of~\cite{Lindblom:2005qh}. The choices of damping parameters for this project are thus~$\alpha \gamma_0 = 2$, $ \alpha \gamma_2 = 10$, $\gamma_1 = -1$, 
and~$\gamma_4=\gamma_5=\frac{1}{2}$ for the case of~$m=1$ initial data, and~$\alpha \gamma_0 = 2$, $ \alpha \gamma_2 =  8$, $\gamma_1 = -1$, 
and $\gamma_4=\gamma_5=\frac{1}{2}$ for the case of~$m=2$. See~\cite{cors2023formulation} for details on the presence of the lapse function,~$\alpha$, in these parameters.

We used the standard NR notation for the lapse~$\alpha$, shift~$\beta^i$ and spatial metric~$\gamma_{ij}$. At the outer boundary, we impose constraint-preserving and radiation-controlling boundary conditions. These conditions eliminate incoming gravitational wave modes by setting the Weyl scalar~$\Psi_0$ to zero, and similarly enforce vanishing values for all incoming characteristic variables of the constraint subsystem. Further details can be found in~\cite{Rinne:2006vv,Ruiz:2007hg,hilditch2016pseudospectral}.

\subsubsection{Matter}
\label{subsubsection:matter}

The second-order equations for the matter content in the spacetime, the complex field~$\psi$, are also reduced to a first-order system of equations. Thus, we introduce the time and space derivatives $\Pi = n^a\partial_a \psi$, $\Phi_i=\partial_i \psi$ as reduction variables. The spatial reduction constraint~$S_i :=\partial_i \psi-\Phi_i$  is used in the equations in order to control the damping. The equations of motion of the complex scalar field are split into real and imaginary parts and take the general form shown below
\begin{align}
    \partial_t \psi={}&\alpha \Pi + \beta^i \Phi_i\,, \\
    \partial_t \Phi_i ={}&\Pi \partial_i\alpha+ \alpha\partial_i \Pi+\sigma\alpha S_i+\Phi_j\partial_i\beta^j+\beta^j\partial_j \Phi_i\,,\\
    \partial_t \Pi={}&  \beta^i \partial_i \Pi +\alpha\Pi K +\sigma \beta^i S_i \nonumber\\
    &+ \gamma^{ij}\big(\Phi_j\partial_i\alpha+\alpha\partial_i \Phi_j -\alpha {}^{(3)}\Gamma^k{}_{ij}\Phi_k\big)\,.
\end{align}
We stress that~$\Phi_i$ (with indices) is the spatial reduction variable of the field, while small~$\phi$ without indices denotes the azimuthal angle.

We evolve the fields~$\Phi^\mathrm{Re}_i$, $\Phi^\mathrm{Im}_i$, 
$\Pi^\mathrm{Re}$, $\Pi^\mathrm{Im}$, $\psi^\mathrm{Re}$, and~$\psi^\mathrm{Im}$. Here, $\partial_i \alpha$ is computed from the evolved variables as~$\partial_i \alpha = -\tfrac{1}{2} \alpha \, n^a n^b \Phi_{iab}$. Likewise, the trace of the extrinsic curvature $K$ and the spatial Christoffel symbols~${}^{(3)}\Gamma^k{}_{ij}$ are obtained by assuming the reduction constraints are satisfied and forming suitable combinations of the reduction variables. Our implementation follows the approach used in earlier complex scalar field simulations~\cite{Atteneder:2023pge,MaroudaCors2024}, with the potential term suppressed. This scheme is itself based on our earlier implementation of the real scalar field in~\cite{Bhattacharyya:2021dti,cors2023formulation}, which has already been extensively tested. For the scalar field reduction constraint, we set the damping parameter to~$\alpha \sigma = 10$ for the $m=1$ families and~$\alpha \sigma = 8$ for the~$m=2$ families, in direct analogy with our treatment of~$\gamma_2$. At the outer boundary, 
we impose reduction constraint preserving conditions together with Sommerfeld-type conditions on the incoming physical characteristic variables.

\begin{table*}[t!]
   \label{tbl:initialdata}
    \begin{ruledtabular}
    \begin{tabular}{ccccccccccccc}
    \textbf{Family} & m & $\varepsilon^{\mathrm{Re}}$ &\textbf{$a^\textrm{Re}$} & \textbf{$r_0^\textrm{Re}$ }& \textbf{$s_z^\textrm{Re}$}& \textbf{$s_\rho^\textrm{Re}$} & $\varepsilon^{\mathrm{Im}}$ &\textbf{$a^\textrm{Im}$} &  \textbf{$r_0^\textrm{Im}$}  & \textbf{$s_z^\textrm{Im}$} & \textbf{$s_\rho^\textrm{Im}$} & $w$\\  \hline
     \textbf{I}& 1 &  1 & tuned  & 0.6 &1.0  &  1.0 & -1 & $a^\textrm{Re}$ & 0.6 &1.0 &1.0 & 0.1\\
      \textbf{II}& 1 & 1 & $3a^\textrm{Im}$  & 0.6 &1.0 &1.0 & -1 &tuned & 0.6 &1.0 &1.0 & 0.1\\ 
      \textbf{III}&2& 1 &tuned & 0.6 & 1.0 &1.0 & -1 &  $a^\textrm{Re}$ & 0.6 &1.0 &1.0 & 0.1\\  
      \textbf{IV}&2&1& tuned & 0.8 &1.0 & 1.0 & -1 & $2a^\textrm{Re}$ & 0.8 &1.0 &1.0 & 0.2
    \end{tabular}
    \end{ruledtabular}
    \caption{Specification of the values appearing in Eqs.~(\ref{initial_data_1}-\ref{initial_data_4}) for each family of initial data. Note that the first two families refer to the case of $m=1$ while the last two refer to the case of $m=2$.}
\end{table*}

\subsubsection{Initial data}
\label{subsubsection:IDsolver}

The initial data containing angular momentum follows the form given in~\cite{choptuik2004critical}, adapted to our evolved variables, $\left\{\psi_{\textrm{Re}},\psi_{\textrm{Im}},\Pi_{\textrm{Re}},\Pi_{\textrm{Im}}\right\}$. However, the main difference is that in our case, a symmetric pulse about the origin has been added to ensure regularity at the center, since our pseudospectral code absolutely requires smooth data to perform accurately. The initial data of~\cite{choptuik2004critical} are not regular, which renders their reproduction impossible using pseudospectral methods without more care. The analytic expressions we use for our initial data solver are as follows

\begin{align}
\label{initial_data_1}
   \psi^{\mathrm{Re}} 
   ={}& a^{\mathrm{Re}} \, \Re\!\big[(x+iy)^m\big] \notag\\
   \notag & \Bigg[
\exp\Bigg(-\frac{\Big(\sqrt{\left(\frac{z}{s_z^{\mathrm{Re}}}\right)^2 + \left(\frac{\rho}{s_\rho^{\mathrm{Re}}}\right)^2} - r_0^{\mathrm{Re}}\Big)^2}{w^2}\Bigg) \\
\notag &  \phantom{\Bigg[}+ \exp\Bigg(-\frac{\Big(\sqrt{\left(\frac{z}{s_z^{\mathrm{Re}}}\right)^2 + \left(\frac{\rho}{s_\rho^{\mathrm{Re}}}\right)^2} + r_0^{\mathrm{Re}}\Big)^2}{w^2}\Bigg)\Bigg] \\
 \notag &- a^{\mathrm{Im}} \, \Im\!\big[(x+iy)^m\big] \\
 \notag &  \Bigg[
\exp\Bigg(-\frac{\Big(\sqrt{\left(\frac{z}{s_z^{\mathrm{Im}}}\right)^2 + \left(\frac{\rho}{s_\rho^{\mathrm{Im}}}\right)^2} - r_0^{\mathrm{Im}}\Big)^2}{w^2}\Bigg) \\
 &  \phantom{\Bigg[}+ \exp\Bigg(-\frac{\Big(\sqrt{\left(\frac{z}{s_z^{\mathrm{Im}}}\right)^2 + \left(\frac{\rho}{s_\rho^{\mathrm{Im}}}\right)^2} + r_0^{\mathrm{Im}}\Big)^2}{w^2}\Bigg)\Bigg]\,,
\end{align}
for the real part, and, 
\begin{align}
   \notag \psi^{\mathrm{Im}} 
   ={}& a^{\mathrm{Im}} \, \Re\!\big[(x+iy)^m\big] \\
   \notag & \Bigg[
\exp\Bigg(-\frac{\Big(\sqrt{\left(\frac{z}{s_z^{\mathrm{Im}}}\right)^2 + \left(\frac{\rho}{s_\rho^{\mathrm{Im}}}\right)^2} - r_0^{\mathrm{Im}}\Big)^2}{w^2}\Bigg) \\
\notag & \phantom{\Bigg[}+ \exp\Bigg(-\frac{\Big(\sqrt{\left(\frac{z}{s_z^{\mathrm{Im}}}\right)^2 + \left(\frac{\rho}{s_\rho^{\mathrm{Im}}}\right)^2} + r_0^{\mathrm{Im}}\Big)^2}{w^2}\Bigg)\Bigg]  \\
 \notag &+ a^{\mathrm{Re}} \, \Im\!\big[(x+iy)^m\big] \\ 
\notag & \Bigg[
\exp\Bigg(-\frac{\Big(\sqrt{\left(\frac{z}{s_z^{\mathrm{Re}}}\right)^2 + \left(\frac{\rho}{s_\rho^{\mathrm{Re}}}\right)^2} - r_0^{\mathrm{Re}}\Big)^2}{w^2}\Bigg) \\
 & \phantom{\Bigg[}+ \exp\Bigg(-\frac{\Big(\sqrt{\left(\frac{z}{s_z^{\mathrm{Re}}}\right)^2 + \left(\frac{\rho}{s_\rho^{\mathrm{Re}}}\right)^2} + r_0^{\mathrm{Re}}\Big)^2}{w^2}\Bigg)\Bigg]\,,
\label{initial_data_2}
\end{align}
for the imaginary part. We observe in passing that the these scalar field data consist of pure~$l=|m|$ spherical harmonic modes only. This property is not likely to be propagated in strong-field time evolutions.

The explicit formulas that we use for the time derivatives are
\begin{align}
  \notag \Pi^{\mathrm{Re}} 
  ={}& a^{\mathrm{Re}} \, \varepsilon^{\mathrm{Re}} \, \Re\!\big[(x+iy)^m\big] \\
  \notag & \Bigg[
\exp\Bigg(-\frac{\Big(\sqrt{\left(\frac{z}{s_z^{\mathrm{Re}}}\right)^2 + \left(\frac{\rho}{s_\rho^{\mathrm{Re}}}\right)^2} - r_0^{\mathrm{Re}}\Big)^2}{w^2}\Bigg) \\
\notag & \phantom{\Bigg[}+ \exp\Bigg(-\frac{\Big(\sqrt{\left(\frac{z}{s_z^{\mathrm{Re}}}\right)^2 + \left(\frac{\rho}{s_\rho^{\mathrm{Re}}}\right)^2} + r_0^{\mathrm{Re}}\Big)^2}{w^2}\Bigg)\Bigg] \\
\notag &- a^{\mathrm{Im}} \, \varepsilon^{\mathrm{Im}} \, \Im\!\big[(x+iy)^m\big] \\
\notag & \Bigg[
\exp\Bigg(-\frac{\Big(\sqrt{\left(\frac{z}{s_z^{\mathrm{Im}}}\right)^2 + \left(\frac{\rho}{s_\rho^{\mathrm{Im}}}\right)^2} - r_0^{\mathrm{Im}}\Big)^2}{w^2}\Bigg) \\
& \phantom{\Bigg[}+ \exp\Bigg(-\frac{\Big(\sqrt{\left(\frac{z}{s_z^{\mathrm{Im}}}\right)^2 + \left(\frac{\rho}{s_\rho^{\mathrm{Im}}}\right)^2} + r_0^{\mathrm{Im}}\Big)^2}{w^2}\Bigg)\Bigg]\,,\label{initial_data_3}
\end{align}
for the real part, and, 
\begin{align}
  \notag \Pi^{\mathrm{Im}} 
  ={}& a^{\mathrm{Im}} \, \varepsilon^{\mathrm{Im}} \, \Re\!\big[(x+iy)^m\big] \\
  \notag & \Bigg[
\exp\Bigg(-\frac{\Big(\sqrt{\left(\frac{z}{s_z^{\mathrm{Im}}}\right)^2 + \left(\frac{\rho}{s_\rho^{\mathrm{Im}}}\right)^2} - r_0^{\mathrm{Im}}\Big)^2}{w^2}\Bigg) \\
\notag & \phantom{\Bigg[}+ \exp\Bigg(-\frac{\Big(\sqrt{\left(\frac{z}{s_z^{\mathrm{Im}}}\right)^2 + \left(\frac{\rho}{s_\rho^{\mathrm{Im}}}\right)^2} + r_0^{\mathrm{Im}}\Big)^2}{w^2}\Bigg)\Bigg]  \\ 
\notag &+ a^{\mathrm{Re}} \, \varepsilon^{\mathrm{Re}} \, \Im\!\big[(x+iy)^m\big] \\
\notag & \Bigg[
\exp\Bigg(-\frac{\Big(\sqrt{\left(\frac{z}{s_z^{\mathrm{Re}}}\right)^2 + \left(\frac{\rho}{s_\rho^{\mathrm{Re}}}\right)^2} - r_0^{\mathrm{Re}}\Big)^2}{w^2}\Bigg) \\
& \phantom{\Bigg[}+\exp\Bigg(-\frac{\Big(\sqrt{\left(\frac{z}{s_z^{\mathrm{Re}}}\right)^2 + \left(\frac{\rho}{s_\rho^{\mathrm{Re}}}\right)^2} + r_0^{\mathrm{Re}}\Big)^2}{w^2}\Bigg)\Bigg]\,,
\label{initial_data_4}
\end{align}
for the imaginary part. The specific values for the set of parameters $\left\{m, \varepsilon^{\mathrm{Re/Im}}, a^\textrm{Re/Im}, r_0^\textrm{Re/Im}, s_z^\textrm{Re/Im}, s_\rho^\textrm{Re/Im}, w \right\}$ for each family are shown in Table~\ref{tbl:initialdata}. As outlined next, the Hamiltonian and momentum constraints are solved with these functions as given data, producing initial data for which the residual violations are at most of order~$O(10^{-7})$ initially and throughout most of the evolution. 

We solve for conformally flat initial data, $\gamma_{ij}=\psi^4\bar{\gamma}_{ij}=\psi^4\bar{\delta}_{ij}$, with~$\partial_t \bar{\gamma}_{ij}=0$,
together with maximal slicing, $K = 0$, and~$\partial_t K = 0$, using the extended conformal thin-sandwich (XCTS) formulation of the constraint 
equations~\cite{PfeYor03,BauSha10,Tic17}. At the outer boundary, we impose Robin boundary conditions on $\psi_\mathrm{conf}$, $\alpha$, and $\beta^i$, chosen to be compatible with a $1/r$ falloff toward the flat-space values. The initial data for the complex scalar field is described in Sec.~\ref{sec:numerical_results}. The XCTS equations constitute a coupled system of elliptic partial differential equations, which we solve using the hyperbolic relaxation method implemented in \textsc{bamps}~\cite{Ruter:2017iph}. In Table~\ref{tbl:elliptic_parameters} we give an overview of the parameters used for the solution of the elliptic problem.

\begin{table}[t!]
 \centering
 \begin{ruledtabular}
 \begin{tabular}{lllll}
      Parameter & Value \\
      \hline
      \texttt{elliptic.maxiterations} & $10^9$                    \\
      \texttt{elliptic.dtfactor} &   0.25           \\
      \texttt{elliptic.method} & rk4\\
      \texttt{elliptic.startgrid.n.xyz} & 5 \\
      \texttt{elliptic.increment}  &   2  \\
      \texttt{elliptic.rcheck.every}  &  $10^4$  \\
      \texttt{elliptic.rcheck.factor} &  $10^{-3}$  \\
      \texttt{elliptic.XCTS.initialguess} & const   \\
      \texttt{elliptic.XCTS.initialguess.const.amp} &   1.0 1.0 0.0 0.0 0.0  \\
  \end{tabular}
  \end{ruledtabular}
 \label{tbl:elliptic_parameters}
 \caption{
   Summary of the elliptic solver parameters used for all the families studied in this project. 
   Parameter description:
   \texttt{elliptic.maxiterations}: maximum number of iterations to obtain the necessary convergence of the solution,
   \texttt{elliptic.dtfactor}: time step for the method,
   \texttt{elliptic.method}: integration method is 4th order Runge-Kutta,
   \texttt{elliptic.startgrid.n.xyz}: number of points per dimension on the Gauss-Lobatto grid in each cell before any round of mesh refinement,
   \texttt{elliptic.increment}: number of points added in each round of mesh refinement of the relaxation method,
   \texttt{elliptic.rcheck.every}: determines the frequency with which the code checks for whether to change the number of points,
   \texttt{elliptic.rcheck.factor}: determines when the accuracy should be increased and the increment should be applied,
   \texttt{elliptic.XCTS.initialguess}: guess of constant flat metric initially,
   \texttt{elliptic.XCTS.initialguess.const.amp}: initial guess values for the conformal factor, the lapse and the components of the shift vector.}
\end{table}

\subsection{Grid setup}
\label{subsec:grid}

In \textsc{bamps}, the fundamental computational units are cubic cells, 
each of which solves its own Initial Boundary Value Problem (IBVP). Communication between neighboring cells is handled via the penalty method as described in~\cite{hilditch2016pseudospectral}. If the neighboring grid cells have different resolutions, an additional interpolation step is required for this data exchange \cite{renkhoff2023adaptive}. Within each grid, the numerical solution is represented by a nodal pseudospectral expansion based on Gauss-Lobatto-Chebyshev collocation points in each dimension, providing an efficient approximation of spatial derivatives throughout the evolution. Temporal discretization is carried out using the method of lines with a fourth-order Runge-Kutta (RK4) scheme.

The numerical domain representing spatial slices in \textsc{bamps} consists of patches of these local cubic grids, which are organized into three main regions: a central cube covering the strong-field region, a spherical outer shell, and an intermediate cubed-sphere shell that ensures a smooth transition between the two, see Fig.~1 of~\cite{hilditch2016pseudospectral} and Fig.~1 of~\cite{cors2023formulation}. The size of the central cube, the transition shell, and the outer sphere radius are chosen individually for each type of initial data so that the strong-field region lies entirely within the central cube. This is preferred in order to avoid errors coming from the boundaries between the patches, while solving the constraints with the elliptic solver. Further details of the grid configurations are provided in Table~\ref{tbl:grid_parameters}.

\begin{table}[t!]
 \centering
 \begin{ruledtabular}
 \begin{tabular}{lllll}
      Parameter & Setup 1 & Setup 2  \\
      \hline
      \texttt{grid.cube.max}           & 5     &  5     \\
      \texttt{grid.sub.xyz}            & 50    &  80   \\
      \texttt{grid.cubedsphere.max.x}  & 13   &  13    \\
      \texttt{grid.cubedsphere.sub.x}  & 16   &  16   \\
      \texttt{grid.sphere.max.x}       & 25   &  25   \\
      \texttt{grid.sphere.sub.x}       & 24   &  24     \\
      \texttt{grid.dtfactor}           &0.25  &  0.25  \\
      \texttt{grid.cartoon}            & x    &  x   \\
      \texttt{grid.reflect}            & z    &  z   \\
      \texttt{grid.n.xyz}              & [5,25] & [5,19] \\
  \end{tabular}
  \end{ruledtabular}
 \label{tbl:grid_parameters}
 \caption{
   Summary of the grid setup used for all the families presented in this work. Setup 1: families I, II with m=1, Setup 2: families III, IV with m=2.
   Parameter description:
   \texttt{grid.cube.max}: 1/2 side length of inner cube,
   \texttt{grid.sub.xyz}: number of subdivisions in inner cube,
   \texttt{grid.cubedsphere.max.x}: outer radius of cube-to-sphere patch,
   \texttt{grid.cubedsphere.sub.x}: number of radial subdivisions in cube-to-sphere patch,
   \texttt{grid.sphere.max.x}: radius of spherical outermost boundary,
   \texttt{grid.sphere.sub.x}: number of radial subdivisions in sphere patch,
   \texttt{grid.dtfactor}: Courant-Friedrichs-Lewy (CFL) factor,
   \texttt{grid.cartoon}: the remaining dimensions after applying the cartoon method,
   \texttt{grid.reflect}: reflection symmetry across $z=0$ plane,
   \texttt{grid.n.xyz}: number of points per grid and per dimension, which increase in steps of two in our use of $p$-refinement.}
\end{table}

\subsection{\texorpdfstring{$hp$}{hp}-refinement}
\label{subsec:hprefinement}

The spacetimes we investigate demand exceptionally high resolution, especially in the strong-field region. More specifically, solutions near the threshold of collapse are expected to form structure on ever smaller spacetime scales. In the usual picture of critical collapse, they approach a DSS spacetime, defined in Eq.~\eqref{eq:DSS1} and \eqref{eq:DSS2}.

When a family of solutions approaches a DSS critical spacetime, the period~$\Delta$ appears observationally as the temporal separation between successive, self-similar features, commonly referred to as \emph{echoes}, in a suitably chosen coordinate system. Each echo represents a rescaling of the physical geometry, with the characteristic length scale shrinking by a factor of $e^{\Delta}$. For the case of a spherically symmetric, massless, real scalar field, numerical studies show $e^{\Delta} \approx 31$ in the strong-field region when the system is tuned to the threshold of collapse within a generic smooth one-parameter family of initial data. In the present work, we find that the numbers are~$e^{\Delta} \approx 1.5$ for the~$m=1$ families and~$e^{\Delta} \approx 1.1$ for the~$m=2$ families, which favors the appearance of more echoes before reaching the limits of machine precision.

Our adaptive mesh refinement scheme ensures the necessary accuracy for studying the threshold of collapse in the set of solutions under consideration. Two complementary strategies are employed: reducing the physical size of grids ($h$-refinement) and increasing the polynomial order of the spectral expansion within each grid, thereby adding collocation points ($p$-refinement). In the present work, we apply the $hp$-adaptive mesh refinement scheme introduced in~\cite{renkhoff2023adaptive}. Here, $h$-refinement is driven by smoothness indicators, while $p$-refinement is controlled by an error estimate derived from the spectral truncation order in each grid.

We allow up to~$15$ successive $h$-refinement levels, halving the grid spacing at every step, and up to 31 collocation points per dimension per grid for $p$-refinement, increased in steps of two. Refinement or de-refinement in either~$h$ or~$p$ occurs whenever the corresponding indicator~$\epsilon$ leaves its admissible range~$[\epsilon_\mathrm{min}, \epsilon_\mathrm{max}]$. The smoothness interval of our choice for the $h$-refinement is $[0.008, 0.04]$. For $p$-refinement, the target error is uniformly set to~$[10^{-12}, 10^{-10}]$. Following~\cite{renkhoff2023adaptive}, these indicators are applied 
to both the metric and the scalar field variables. Time evolution is parallelized using the Message Passing Interface (MPI), ensuring efficient distribution of work across computational resources. All simulations were produced using between 768 and 4800~CPU cores on SuperMUC-NG, depending on the required resolution for each family, for a maximum of 5 hours each. In total we used around~$6$ million core hours for this work, which translates to $64000~\textrm{kWh}$. 

\subsection{The m-cartoon method} 
\label{subsec:cartoom}

Symmetry reduction for evolving axisymmetric spacetimes with a~$2+1$  computational domain instead of the full~$3+1$ dimensions is achieved in \textsc{bamps} using the cartoon method. This approach exploits the existence of a Killing vector associated with invariance under azimuthal rotations to
construct a prescription for calculating~$y$-derivatives. As a result, one may evolve solely on the~$xz$-plane (or equivalently the~$y=0$ plane) and reconstruct the third spatial dimension in post-processing if needed. In our case, while the spacetime is axisymmetric, the complex scalar field itself is not. We therefore introduce what we call the \emph{m-cartoon} method, a generalization to the cartoon method, which follows the same computational strategy, see~\cite{Alcubierre:1999ab,Pretorius:2004jg,hilditch2016pseudospectral}, but adapted
to the new relaxed symmetries of the problem. More precisely, the prescription of the~$y$-derivatives of the field in the~$y=0$ plane need to be adapted to account for the~$\phi$-dependence of the field, completely captured by its winding number~$m$. In the name `m-cartoon', the letter `m' is taken from the specific mode~$m$ imposed on the field throughout the entire evolution (see Eq.~\eqref{eq:ansatz}).

As stated above, we place an ansatz for a single mode~$\psi = e^{im\phi} \psi_m(t,\rho,z)$, on the scalar field. Recalling~\eqref{eq:Killing} and restricting to the~$y=0$ plane, for $x \neq 0$, we thus have,
\begin{align}
    \label{eq:mcartoon1}
    \partial_y \psi = i m \frac{\psi}{x}\,,
\end{align}
for the~$y$-derivatives. 
so that on the~$z$-axis, where~$x\to0$, we have to prescribe the limit separately
\begin{align}
    \label{eq:mcartoon2}
    \lim_{x\to0} \partial_y \psi = i m \partial_x \psi\,.
\end{align}
These two prescriptions~\eqref{eq:mcartoon1}, \eqref{eq:mcartoon2} for finding~$y$-derivatives at the interior of the domain and on the axis $x=0$ respectively apply to the complex scalar field and its time derivative $\Pi = n^\mu \partial_\mu \psi$ as well, which is also a scalar.

For covectors, such as the spatial derivatives of the field, $\Phi_i = \partial_i \psi$, we work accordingly
\begin{align}
    \Phi_i = \partial_i \psi &= \partial_i\left(e^{im\phi}\psi_m(t,\rho,z)\right) \\
    \notag &= e^{im\phi} \partial_i \psi_m(t,\rho,z) + im \left(\partial_i \phi\right) \underbrace{e^{im\phi}\psi_m(t,\rho,z)}_\psi
\end{align}
the~$\phi$-derivative is then calculated as 
\begin{align}
\label{3vector}
  \begin{split}
    \partial_\phi \Phi_i ={}& im e^{im\phi} \partial_i \psi_m(t,\rho,z) +  e^{im\phi} \partial_\phi \partial_i \psi_m(t,\rho,z) \\
     &+ im\left(\partial_\phi \partial_i \phi\right) \overbrace{e^{im\phi} \psi_m(t,\rho,z)}^{\psi}
    \\
     &
    + im\left(\partial_i \phi\right) \overbrace{\partial_\phi \left(e^{im\phi}\psi_m(t,\rho,z)\right)}^{im\psi}\,.
  \end{split}
\end{align}
We will now prepare the quantities that appear in the right-hand side of the previous expression written in Cartesian coordinates, since \textsc{bamps} works in these. The Cartesian derivatives of the azimuthal angle are
\begin{align}
    \partial_i \phi =\partial_i \arctan\left(\frac{y}{x}\right)=\partial_i\left(\frac{y}{x}\right)\frac{x^2}{\rho^2}\,,
\end{align}
where $\rho ^2= x^2+y ^2$ and $y=\rho \cos\phi$, $x=\rho \sin\phi$, then for $x>0$ we have,
\begin{align}
    \partial_i \phi = \left(-\frac{y}{\rho ^2},\frac{x}{\rho ^2},0\right)=\left(-\frac{\sin\phi}{\rho },\frac{\cos\phi}{\rho },0\right)\,.
\end{align}
Subsequently, we find,
\begin{align}
    \partial_\phi \partial_i \phi = \left(-\frac{\cos\phi}{\rho },-\frac{\sin\phi}{\rho },0\right) = \left(-\frac{x}{\rho ^2},-\frac{y}{\rho ^2},0\right)\,.
\end{align}
Additionally, we calculate, 
\begin{align}
    \partial_i \psi_m &= \left(\partial_x\rho ~ \partial_\rho \psi_m, \partial_y \rho  ~ \partial_\rho \psi_m, \partial_z \psi_m\right) \\
    \notag &= \left(\frac{x}{\rho}\partial_\rho \psi_m, \frac{y}{\rho}\partial_\rho \psi_m, \partial_z \psi_m\right)\,,
\end{align}
where here and henceforth, we suppress the arguments of~$\psi_m\left(t,\rho,z\right)$.
Finally, we will need
\begin{align}
    \partial_\phi \partial_i \psi_m = \left(-\frac{y}{\rho}\partial_\rho \psi_m, \frac{x}{\rho}\partial_\rho \psi_m, 0\right)\,.
\end{align}
In order to calculate the~$y$-derivatives of the 3-vectors we now proceed to rewrite Eq.~\eqref{3vector} as follows
\begin{align}
    \left(x\partial_y - y \partial_x\right) \Phi_i &= im e^{im\phi} \partial_i \psi_m +  e^{im\phi} \partial_\phi \partial_i \psi_m \notag\\
    &+ im\left(\partial_\phi \partial_i \phi\right)\psi-m^2\left(\partial_i \phi\right) \psi \equiv V_i  \,,
\end{align}
where we gave a name $V_i$ to the whole right-hand side of the previous expression. Then, similarly to the case of scalars we get,
\begin{align}
    \partial_y \Phi_i = \frac{y}{x}\partial_x \Phi_i + \frac{V_i}{x}\,.
\end{align}
Substituting the components of the expression $V_i$, and restricting to $y=0$, for $x\neq 0$, we have
\begin{align}
  \begin{split}
    \partial_y \Phi_i ={}& im \left(\frac{1}{\rho}\partial_\rho\psi,0, \frac{\partial_z \psi}{x}\right) + \left(0,\frac{1}{\rho}\partial_\rho\psi,0\right) \\
    &+ im \left(-\frac{1}{\rho^2},0,0\right)\psi
    - m^2 \left(0,\frac{1}{\rho^2},0\right)\psi\,.
  \end{split}
\end{align}
To replace $\partial_\rho \psi$, we use the fact that 
\begin{align}
    \partial_x \psi &= \frac{\partial \rho}{\partial x}\partial_\rho \psi + \frac{\partial \phi}{\partial x}\partial_\phi \psi 
    = \frac{x}{\rho}\partial_\rho \psi - \frac{y}{\rho^2}\partial_\phi \psi\,,
\end{align}
from which, when~$y=0$, for $x\neq 0$, we have
\begin{align}
\partial_\rho \psi &= \partial_x \psi \,,
\end{align}
in order to find
\begin{align}
  \begin{split}
    \partial_y \Phi_i ={}& im \left(\frac{1}{x}\partial_x\psi,0, \frac{\partial_z \psi}{x}\right) + \left(0,\frac{1}{x}\partial_x\psi,0\right) \\
    &+ im \left(-\frac{1}{x^2},0,0\right)\psi 
    - m^2 \left(0,\frac{1}{x^2},0\right)\psi\,.
  \end{split}
\end{align}
Taking now the limit of $x\to 0$ for each of the components leads to
\begin{align}
    \partial_y \Phi_x &= im \frac{1}{x^2}\left(x \partial_x \psi -\psi\right)  \notag\\
    \notag \lim_{x\to0} \partial_y \Phi_x 
    &= \lim_{x\to0} \frac{im}{2x}\left(\partial_x\psi +x\partial^2_x \psi -\partial_x \psi\right) \\
    \notag &= \frac{im}{2}\partial^2_x \psi = \frac{im}{2}\partial_x\Phi_x\\
    \notag \partial_y \Phi_y &= \frac{1}{x^2}\left(x \partial_x \psi - m^2 \psi\right) \\
     \notag \lim_{x\to0} \partial_y \Phi_y 
     &= \lim_{x\to0} \frac{1}{2x}\left(\partial_x\psi+x\partial^2_x\psi-m^2\partial_x\psi\right) \\
    \notag &= \frac{1}{2}\left(\partial^2_x\psi +\partial^2_x\psi+x\partial^3_x\psi-m^2\partial^2_x\psi\right) \\
    \notag &= \left(1-\frac{1}{2}m^2\right)\partial^2_x\psi = \left(1-\frac{1}{2}m^2\right)\partial_x \Phi_x \\
    \notag \partial_y\Phi_z &= im \frac{\partial_z\psi}{x} \\
    \lim_{x\to0} \partial_y \Phi_z &= im \partial_x \partial_z \psi =  im \partial_x \Phi_z \,,
\end{align}
where second derivatives of~$\psi$ have been replaced using the appropriate reduction variables in the code. 
These relations give us a prescription for the calculation of the~$y$-derivatives of the spatial reduction variables of the complex scalar field on the~$y=0$ plain and in the limit of the~$x=0$ axis, meaning the~$z$-axis. The way this information is passed into the code is by splitting the expressions into real and imaginary parts, since the code solves for them separately. A final detail required for the numerical implementation concerns the parity conditions at the origin: along the~$x$-direction, the field exhibits even parity for even~$m$ and odd parity for odd~$m$.

\subsection{Classification of spacetimes with twist}
\label{subsec:phase_space_search}

We use the quasilocal notion of an apparent horizon in order to determine the existence of a black hole in our spacetime evolutions. A foliation-independent approach could instead be obtained by tracing null rays to locate the event horizon, though this would require substantially more post-processing of the full spacetime data. On the other hand, apparent horizon finders allow us to search on each numerical slice separately and possibly detect the ``outermost marginally trapped surface'' (MOTS) that may lie on a Cauchy slice, $\Sigma$, of the spacetime. A MOTS is defined as a compact 2-dimensional submanifold of~$\Sigma$, $S \subseteq \Sigma$, where the expansion $\Theta$ of all outgoing null geodesics is zero. Following from Penrose's singularity theorems~\cite{Penrose:1964wq}, the assumption of global hyperbolicity, a choice of energy condition, and the existence of a trapped surface signals the birth of a singular spacetime, meaning future geodesic incompleteness. Assuming that the cosmic censorship conjecture is true, then the apparent horizon will lie in the interior of the event horizon, and in the case of stationary spacetimes those two notions should be identical. For the aforementioned reasons, we select the apparent horizon as a diagnostic for the presence of a black hole since it is numerically more efficient while maintaining a relationship to the existence of an event horizon.

Let an arbitrary 2d surface with normal vector~$s^i$, then the expansion of null geodesics on that surface is 
\begin{align}
    \Theta \equiv D_i s^i + K_{ij}s^i s^j-K \, ,
\end{align}
where~$K_{ij}$ is the extrinsic curvature, $K$ its trace, and~$D_i$ is the induced spatial covariant derivative. The expansion of outgoing null geodesics is everywhere zero on a MOTS. In the case of axisymmetry, in order to find the locus of points that satisfy this condition one has to find solutions to a differential equation.

The numerical tool we are using is \textsc{AHloc3d}~\cite{ahloc3d}, which starts with the assumption of an arbitrary star-shaped 2d surface on each slice and proceeds with a flow method, following up with a Newton-Raphson method in the end once a good enough candidate surface is found. In the case of axisymmetry, our earlier iteration of this tool assumed vanishing twist, so we generalized it in order to be able to detect horizons with angular momentum, and to perform a surface integral in order to calculate the angular momentum on the horizon as in~\cite{PhysRevLett.89.261101}.  A dynamical horizon~$H$ is a 3-manifold which is foliated by marginally trapped 2-spheres, $S$. A quasilocal measure of angular momentum on the horizon~$S$ (compare to Eq.~\eqref{eq:appJAH}) is given by
\begin{align}
\label{eq:JAH}
    J^{\phi}_{\textrm{AH}}= -\frac{1}{8\pi} \oint_{S} K_{ab}\xi^a \hat{r}^b d^2V\,,
\end{align}
where~$\hat{r}^b$ is the unit space-like vector orthogonal to~$S$ and tangent to~$H$. We tested our generalization of the \textsc{AHloc3d} code using the Kerr spacetime, where we input the metric and extrinsic curvature in Kerr-Schild coordinates for various values of the parameter~$a$, which is related to angular momentum of the Kerr black hole, $J_{AH}=a_{\textrm{Kerr}}M$. In this setup, we were able to locate the horizon numerically and analytically confirming that they coincide, and measuring~$J_{AH}$ accurately on the horizon. 

\textsc{AHloc3d} is a post-processing tool that demands costly output from our evolution code, both in terms of runtime and storage. At each bisection level, our strategy is to carry out a series of runs, some of which clearly disperse to infinity (subcritical), while others blow up (supercritical candidates). The latter cases were repeated with binary output (\texttt{output.ah}) enabled only during the final stages of the evolution, starting from the time when horizon formation could be anticipated from the curvature invariants. This approach required running each simulation twice but allowed us to restrict large binary output files to the relevant timesteps. This binary output serves as input for \textsc{AHloc3d}, which subsequently provides the relevant classification. It is worth noting that the version of the finder adapted to data with twist, which we developed in this work, does not significantly affect the computational time required for apparent horizon searches.

In this paper, we were able to classify the spacetimes solely with the use of \textsc{AHloc3d}~\cite{ahloc3d} all the way up to~$15$ decimal digits, as opposed to our previous work~\cite{MaroudaCors2024}, where the finder was limiting our search in some cases above some degree of tuning. We believe this ease in the supercritical classification is also due to the fact that we do not see any bifurcations nor departures from the spherical critical collapse picture (see the discussion below, Sec. ~\ref{sec:numerical_results}), which tend to involve horizon shapes that can be challenging for~\textsc{AHloc3d} to detect~\cite{SuarezFernandez:2022hyx}. Apparent horizon finding in this work is a more trustworthy criterion for the accurate classification of black hole spacetimes. Therefore, the results in this paper are more robust and allow us to study both the supercritical and subcritical sides of the problem.

For a comprehensive review of horizon-finding techniques in numerical relativity, see~\cite{Thornburg:2006zb}. Examples of horizon searches beyond the star-shaped assumption can be found in~\cite{Pook-Kolb:2018igu}. More recent overviews are available in the form of PhD theses~\cite{Chu12,Poo20}.

\section{Numerical Results}
\label{sec:numerical_results}

We now present our numerical results, starting with~$m=1$ evolutions before turning to the~$m=2$ case. We then consider the angular momentum and gravitational wave content near the threshold.

\subsection{m=1 families}
\label{subsec:mequals1}

\begin{figure}[t!]
    \centering
    \includegraphics[width=\linewidth]{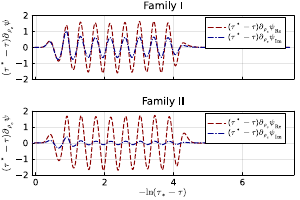}
    \caption{Scaled derivatives of the scalar field at the center for the best tuned subcritical evolutions of Family~I (top) and Family~II (bottom). 
The quantity $(\tau_*-\tau)\,\partial_{\rho_c} \psi$ is plotted against slow time $-\ln(\tau_*-\tau)$. Both the real part (red dashed) and the imaginary part (blue dash-dotted) display oscillatory behavior, with the real component showing larger amplitude oscillations. While the oscillation period seems to be universal across families, the curves themselves are not. Looking at curvature scalars the spacetime metric itself is universal at the threshold.
}\label{fig:m1_central}
\end{figure}

As discussed above, the spacetimes of~\cite{choptuik2004critical} cannot be replicated within our framework due to the requirement of smoothness with our pseudospectral approach. Incidentally, the degree of smoothness is expected in general to effect the behavior of near the threshold. We are concerned with the category of smooth solutions. In any case, the slight differences in our families of initial data allow us to examine universality in the~$m=1$ mode. At the origin the field itself is zero, therefore in order to access the leading order non-zero part of the field we plot the first derivative $\partial_{{\rho_c}} \psi$ with the proper radius at the origin, $\rho_c=\sqrt{g_{\phi\phi}} \stackrel{y=0}{=} \rho \sqrt{g_{yy}} = x \sqrt{g_{yy}}$ and rescaled with~$\left(\tau^*-\tau\right)$, as defined in the introduction.
Figure~\ref{fig:m1_central} shows the highest subcritical evolution of families~I and~II from Table~\ref{tbl:initialdata} with respect to similarity adapted time, as defined in Eq~\eqref{eq:slowT}.
The curves we obtain differ slightly from those reported in~\cite{choptuik2004critical}, with the only distinction being in the amplitudes of the real and imaginary components. A universal curve that would be the same for all families across different works would be the Kretschmann invariant at the origin or any non-zero gauge independent quantity extracted from the stress-energy tensor, such as $T_{ab}T^{ab}$.

\begin{figure*}[!t]
    \centering    
    \vspace{0cm}
    \includegraphics[scale=0.19]{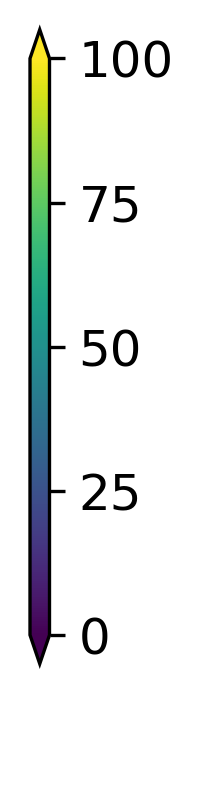} \,
    \includegraphics[scale=0.5]{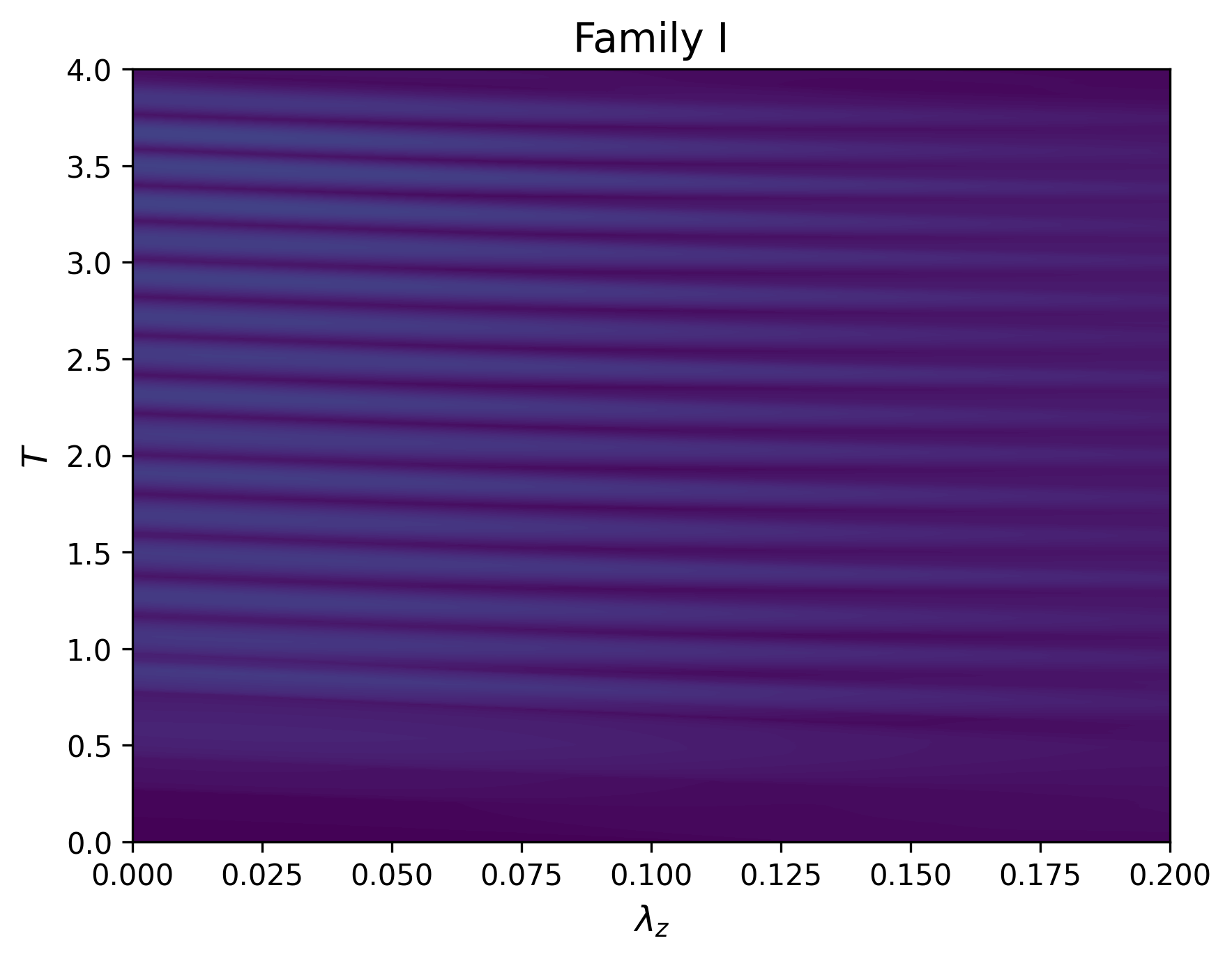} \,
    \includegraphics[scale=0.5]{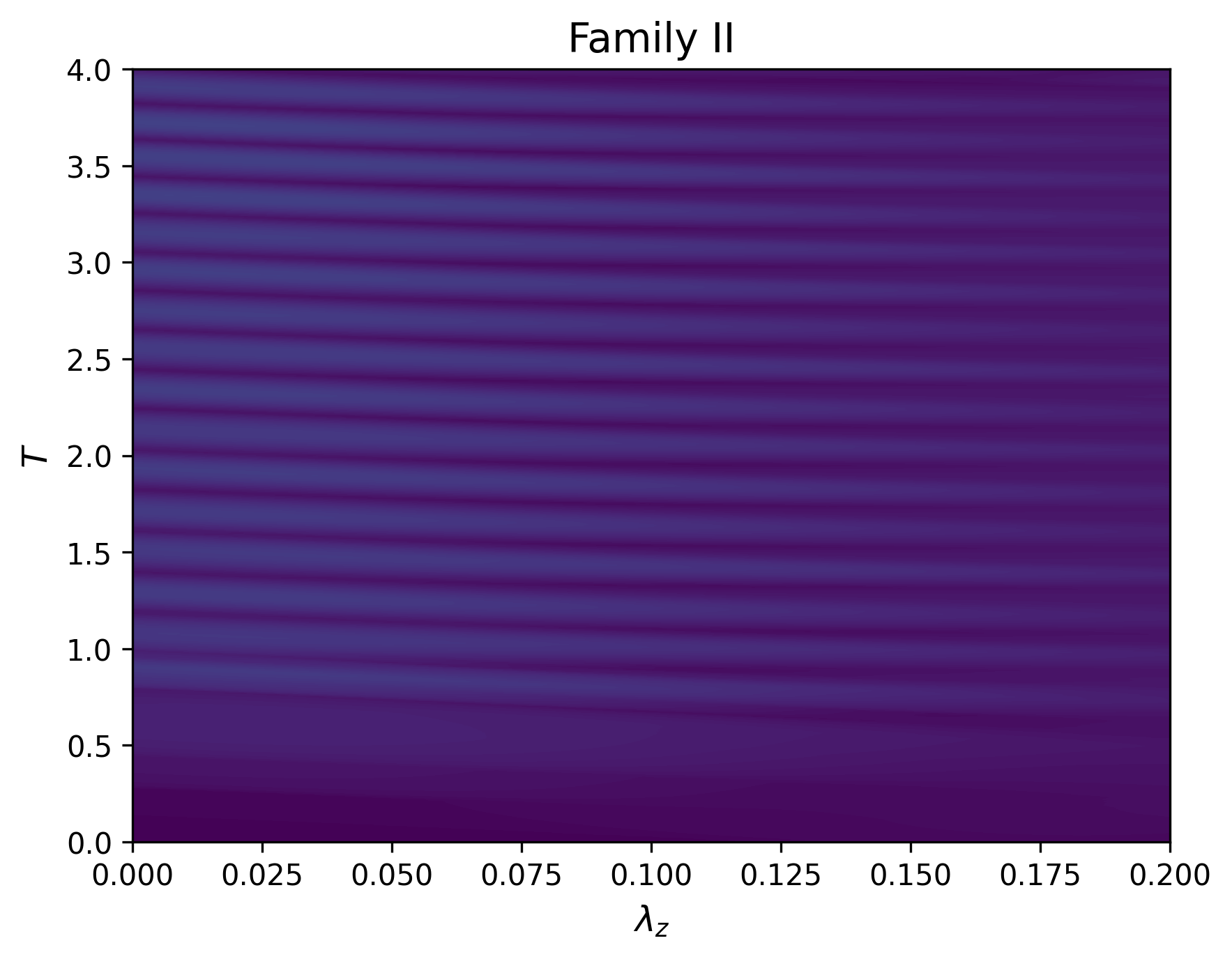} \,
    \caption{Color maps of the normalized Kretschmann invariant~$(\tau_*-\tau) \left|R_{abcd}R^{abcd}\right|^{\frac{1}{4}}$ for our best-tuned subcritical runs of Family I and II initial data setups, all along the symmetry axis in single-null similarity coordinates.} 
\label{fig:m1_Kretsch_nullplots}
\end{figure*} 

In order to define a similarity time coordinate, we select the best-tuned subcritical evolution from each family. The coordinate~$\tau$ is introduced as the proper time at the origin,
\begin{align}
    \tau (t) =\int_{0}^{t} \alpha(t', x^i=0)\,  dt' \, ,
    \label{eq:proper_time}
\end{align}
with~$\alpha$ denoting the lapse function. The accumulation time of DSS~$\tau_*$ is inferred from two independent pairs of zero crossings of the central component of the field shown in the plots~$\left(\tau_n, \tau_{n+1}\right)$ and~$\left(\tau_m, \tau_{m+1}\right)$.
Following the prescription of~\cite{baumgarte2018aspherical}, we evaluate
\begin{align}
    \tau_* = \frac{\tau_n\tau_{m+1}-\tau_{n+1}\tau_{m}}{\tau_n- \tau_{n+1}-\tau_m +\tau_{m+1}}\, ,
    \label{eq:taustar}
\end{align}
and, subsequently, average over multiple pairs of zero crossings. 
The echoing period $\Delta$ is determined using consecutive zero crossings of either the real or imaginary component of the central quantity, one obtains
\begin{align}
    \Delta=2\ln{\frac{\tau_*-\tau_n}{\tau_*-\tau_{n+1}}}\,,
    \label{eq:delta}
\end{align}
since each crossing interval corresponds to half a period, $\Delta/2$. Again, we average over multiple consecutive zero crossings. The value of the period that we find for the~$m=1$ families, $\Delta\simeq0.42$, see Table~\ref{tbl:universalvalues}, is consistent with the results of~\cite{choptuik2004critical}. We note that the period $\Delta$ estimated here corresponds to the oscillation period of each individual curve representing the real and imaginary parts of the central field values, which are not themselves following universal curves (their period is universal, whereas their amplitude is not.

The Kretschmann invariant on the other hand provides better measure of the DSS period, because it exhibits universal behavior.
For instance, in the $m=0$ spherical case discussed in~\cite{MaroudaCors2024},
the period of the Kretschmann invariant was statistically identical to that of the real scalar field.
We confirm that the central value of the Kretschmann invariant against slow time at the origin also follows a universal curve with a period of $\Delta_I\simeq 0.43\pm 0.03$ for both of our $m=1$ families.

Taken together, this strongly suggests that at the threshold, the~$m=1$ families of the complex scalar field coincide qualitatively with those of~\cite{choptuik2004critical}, and that there is universality and DSS. Further support comes from constructing single-null DSS-adapted coordinates, in order to show DSS behavior in an extended region, following~\cite{Baumgarte:2023saw,baumgarte2023critical}. We extend the slow-time variable (now denoted~$T_{\textrm{null}}$) as  
\begin{align}
    T_{\textrm{null}} = -\ln(\tau_*-\tau)\,,
\end{align}
with~$\tau$ and~$\tau_*$ given by Eqs.~\eqref{eq:proper_time} and~\eqref{eq:taustar}. Thus, on the symmetry axis we have~$T_{\textrm{null}}\equiv T$. Outgoing null geodesics are integrated radially from the origin using an affine parameter~$\lambda$, normalized so that~$d\lambda/d\tau=(\tau_*-\tau)^{-1}$ at the center, and labeled by their central value of~$T_{\textrm{null}}$. These curves are computed in post-processing on a uniform grid, and because of interpolation and integration errors, together with the sensitivity of~$T_{\textrm{null}}$ to small shifts in~$\tau_*$, slight adjustment of~$\tau_*$ is required.  

If the spacetime is DSS with accumulation at the center, then the coordinates~$(T_{\textrm{null}},\lambda)$ are symmetry-adapted, and dimensionless scalars should appear periodic in $T_{\textrm{null}}$. We verify this by plotting heat maps of the rescaled Kretschmann scalar, $(\tau_*-\tau) \left|R_{abcd}R^{abcd}\right|^{\frac{1}{4}}$. Figure~\ref{fig:m1_Kretsch_nullplots} shows the results for the best-tuned subcritical evolution family~I and~II. Inspecting by eye, a periodic DSS structure over an extended region away from the origin is observed with period~$\Delta \simeq 0.4$, compatible with the scalar field result.
We observe that in the rotating case such a null-geodesic construction remain possible along the~$z$-axis, but not along the~$x$-axis, where rotation causes photons to rotate around the $z$-axis. 

\begin{table*}[t!]
    \begin{ruledtabular}
    \begin{tabular}{cccccc}
    \textbf{Family} & m & $\Delta $ & $\Delta_I$ & $\gamma_{\textrm{sub}}$ & $\gamma_{\textrm{sup}}$ \\  \hline
     \textbf{I}& 1 &  $0.426 \pm 0.013$ & $0.43 \pm 0.03$  & $0.1105\pm 0.0007$  & $0.098\pm0.003$  \\
      \textbf{II}& 1 & $0.427 \pm 0.012$ & $0.429\pm 0.019$ & $0.1073\pm0.0007$ & $0.113\pm0.004$ \\ 
      \textbf{III}&2& $0.092\pm 0.009$ & $0.092 \pm 0.007$ &$0.0349\pm0.0003$ & $0.031\pm0.003$ \\  
      \textbf{IV}&2& $0.093\pm 0.008$ & $0.088 \pm 0.004$ &$0.0334\pm0.0013$ & $0.0343 \pm 0.0022$
    \end{tabular}
    \end{ruledtabular}
    \caption{Numerical estimation of the parameters that in the Choptuik spherical spacetime appear as universal.
In the first two columns the families are being specified. In the third column the period of DSS as measured by averaging over pairs of zero-crossings of the central values of the fields for each family is presented. The respective errors correspond to the standard deviations of the measured periods. In the fourth column, we calculate the period, $\Delta_I$, of the central value of the Kretschmann invariant, $I$, with respect to slow time, using the maxima of the curve. In the last two columns, we show the regression results from the scaling exponents of the Ricci scalar, Eq.~\eqref{eq:scaling_relation}, at the subcritical side and of the apparent horizon mass, Eq.~\eqref{eq:Mscaling}, at the supercritical side. The errors presented here are the errors resulting from the regression. In our study, the standard picture of universality seems to be maintained for every fixed mode $m>0$.}\label{tbl:universalvalues}
\end{table*} 

\begin{figure}[t!]
    \centering
    \includegraphics[width=\linewidth]{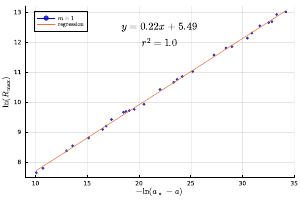}
    \caption{Global maximum of the Ricci scalar, $R_{\text{max}}$, as a function of the inverse of the phase space distance to criticality, $|a-a_*|^{-1}$, for the $m=1$ family I. The data exhibit a clear power-law scaling, with slope $2\gamma_{\textrm{sub}}$. 
Fitting yields a universal critical exponent~$\gamma\simeq 0.11$, in agreement with the results of~\cite{choptuik2004critical}. The regression parameter~$r^2$ quantifies the level of scattering of the data.}\label{fig:m1_sub_scaling}
\end{figure}

Looking into the phase space for subcritical evolutions, we also determine the scaling exponent~$\gamma_{\textrm{sub}}$ associated with the power law
\begin{align}
\label{eq:Rmaxequation}
R_{\text{max}} \propto \left|a-a_*\right|^{-2{\gamma_{\textrm{sub}}}}\,,
\end{align}
by performing a regression fit. As shown in Table~\ref{tbl:universalvalues}, Across both of our~$m=1$ families we obtain~$\gamma_{\textrm{sub}} \simeq 0.11$, in agreement with the values obtained from all families in~\cite{choptuik2004critical}. In Figure~\ref{fig:m1_sub_scaling}, one can see the subcritical power law scaling for family I.

\begin{figure}[t!]
    \centering
    \includegraphics[width=\linewidth]{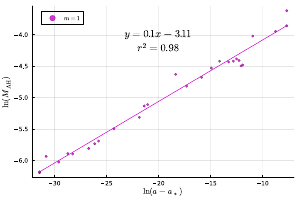}
    \caption{Mass of the first apparent horizon that appears in each simulation, $M_{\text{AH}}$, against the phase space distance to the threshold solution, $|a-a_*|$, for the $m=1$ family~I. The mass scales with an exponent~$\gamma_\textrm{sup}\simeq0.1$ the same universal number found on the subcritical side.}
    \label{fig:m1_sup_scaling}
\end{figure}

For supercritical evolutions, we plot the relation between the mass of the first apparent horizon that appears with respect to the distance in phase space from the critical regime, $|a-a_*|$. 
When the standard picture of critical collapse holds, the exponent for the scaling
law of this mass, given by
\begin{align}
\label{eq:Mscaling}
    M_{\textrm{AH}}\propto (a-a_*)^{\gamma_{\textrm{sup}}}\,,
\end{align}
is expected to be the same scaling exponent we see on the subcritical side, 
$\gamma_{\textrm{sup}}=\gamma_{\textrm{sub}}$.
As illustrated in Fig.~\ref{fig:m1_sup_scaling} for family~I and additionally summarized in Table~\ref{tbl:universalvalues} for family~II, the scaling exponent on the supercritical side of the critical regime is indeed~$\gamma_{\textrm{sup}} \simeq 0.1$, demonstrating the universality of power-law scaling exponents for quantities with dimensions of length in near-threshold solutions with~$m=1$. Upon closer inspection the error margins reported in Table~\ref{tbl:universalvalues} seem to indicate a discrepancy between the super- and subcritical scaling exponents, 
but this can be explained by the fact that the error margins do not account for the systematic error due to finite resolution of the simulation and the foliation dependence of the apparent horizon.

\begin{figure}[t!]
    \centering
    \includegraphics[width=\linewidth]{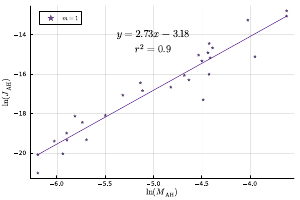}
    \caption{Angular momentum versus mass at horizon formation in near-threshold $m=1$ evolutions of family I.  Each data point shows the angular momentum $J_{\mathrm{AH}}$ against the mass $M_{\mathrm{AH}}$ measured at the first apparent horizon of each evolution. Although both $J_{\mathrm{AH}}$ and $M_{\mathrm{AH}}$ are sensitive to gauge and become very small near the threshold — leading to increased scatter — a clear overall trend is evident. Compare this figure with Fig.\,4 in~\cite{choptuik2004critical}.  We also find that the dimensionless spin $\chi_{\mathrm{AH}} = J_{\mathrm{AH}}/M_{\mathrm{AH}}^{2}$ shows a vanishing trend, decreasing to values~$\mathcal{O}\left(10^{-4}\right)$ at our best tuning, revealing that angular momentum is irrelevant at threshold.}
    \label{fig:m1_JAH_MAH}
\end{figure}

Since the spacetime carries angular momentum, it is natural to measure it at the horizon using Eq.~\eqref{eq:JAH}, and to study its scaling properties. This is subtle because of the foliation dependence of the apparent horizon, but is by far the most convenient option numerically. In Fig.~\ref{fig:m1_JAH_MAH}, we plot the apparent horizon angular momentum against the mass at the moment of formation in our near-critical~$m=1$ simulations of family I. While~$J_{\mathrm{AH}}$, as well as~$M_{\mathrm{AH}}$, are gauge-dependent and subject to numerical noise, especially near the threshold where they become vanishingly small, yet the data nonetheless exhibit a coherent decreasing trend of~$J_{\mathrm{AH}}$ as~$M_{\mathrm{AH}}$ approaches the threshold. Moreover, the dimensionless spin~$\chi_{\mathrm{AH}} = J_{\mathrm{AH}}/M_{\mathrm{AH}}^{2}$ systematically decreases toward the threshold, affirming that angular momentum becomes negligible in the critical regime. In our best-tuned data, we have the dimensionless spin is on the order of $10^{-4}$. From a fit to the data we obtain a scaling of $J_{\mathrm{AH}} ~ \propto M_{\mathrm{AH}}^{2.73}$, as shown in Fig.~\ref{fig:m1_JAH_MAH}.
By comparison, in~\cite{choptuik2004critical} (see Fig.\,4 of that paper) two separate scaling regimes were found~--~a steep~$J\propto M^{6}$ regime very close to threshold, and a milder~$J\propto M^{2.2}$ regime farther from it, suggesting an abrupt slope change in the log-log plot. In our results, though scatter is large, possibly due to gauge effects, we do not resolve such a broken slope. Instead the data are consistent with a single power-law form over our explored mass range. Nonetheless the qualitative agreement remains: angular momentum vanishes more rapidly than~$M^{2}$, in line with the interpretation that it is irrelevant at the threshold within this family.

The foundations of critical collapse were laid out by Choptuik's work on the uncharged massless scalar field in~$3+1$ spherical symmetry, and so scaling behavior of the black hole's angular momentum and charge were taken into account only later. Here we follow the modeling of~\cite{PhysRevD.54.7353,PhysRevD.94.084012} to interpret our data. In particular in~\cite{PhysRevD.94.084012} the authors assume self-similar type II critical collapse and suppose that sufficiently small perturbations of the non-rotating critical solution can be described as a combination of two mode solutions, at least one of which is associated with a growing Lyapunov exponent~$\lambda_0$. The second mode is associated with angular momentum at the threshold and may or may not have a Lyapunov exponent~$\lambda_1>0$. They then use this model to examine scaling of the angular momentum as the threshold is approached. The slightly unusual feature in the context of critical collapse is to consider competing, yet relatively suppressed, dynamics of the second mode. In particular the scaling exponent~$\gamma_{J}$ in the power law of~$J_{AH}$ corresponds to the inverse of the Lyapunov exponent~$\lambda_1=\gamma_{J}^{-1}$, of the rotational mode that differs from the unstable mode with~$\lambda_0=\gamma_{M}^{-1}$ (with~$\gamma_{M}=\gamma_{\textrm{sup}}$ in Eq.~\eqref{eq:Mscaling}) that mediates the collapse in the absence of angular momentum. This explains why~$\gamma_{J}$ is not related to~$\gamma_{M}$ through a simple dimensional analysis, as we have seen can be done for the exponents in the power laws of the Ricci and Kretschmann scalars, whose dynamics are governed at leading order by~$\lambda_0$ (since~$\gamma_{\textrm{sub}}=\gamma_{\textrm{sup}}=\gamma_{M}$ we have~$\gamma_{R}=-2\gamma_{M}$ and~$\gamma_{\textrm{Kretschmann}}=-4\gamma_{M}$). According to this model, we interpret that the scaling of the angular momentum in our $m=1$ families may be governed by a second axial subdominant mode.

\subsection{m=2 families}
\label{subsec:mequals2}

\begin{figure}[t!]
    \centering
    \includegraphics[width=\linewidth]{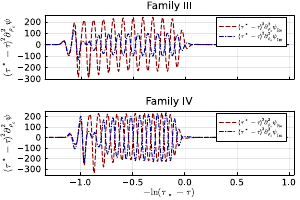}
    \caption{Scaled derivatives of the scalar field at the center for the best-tuned subcritical evolutions in the $m=2$ families~III and IV. The quantity~$(\tau_*-\tau)\,\partial^2_{\rho_c} \psi$ is plotted against slow time~$T=-\ln(\tau_*-\tau)$. Both the real (red dashed) and imaginary (blue dash-dotted) parts display echoing behavior. The shape of the curves of the real and imaginary field components is not universal, but the oscillation period extracted from both families yields $\Delta\simeq 0.09$, significantly smaller than in the $m=1$ case. However, non-zero curvature scalars at the origin are universal. 
}
    \label{fig:m2_central}
\end{figure}

We now investigate~$m=2$ families in order to test the universality of the critical solution across higher angular modes. Since for this data the first derivative of the field at the origin is zero, we access the field through its second derivative, $\partial^2_{\rho_c}\psi$, with $\rho_c=\sqrt{g_{\phi\phi}}$ being the proper radius at the origin, normalizing this time with~$(\tau_* -\tau )^2$. 
Near-critical central values from the evolution of the real and imaginary parts, shown in Fig.~\ref{fig:m2_central}, exhibit echoing behavior analogous to the~$m=1$ case (compare with Fig.~\ref{fig:m1_central}). In this case however the extracted echoing period is~$\Delta\simeq 0.09 \pm 0.01$, noticeably smaller than the~$m=1$ value, and thus signaling a dependence of~$\Delta$ on the angular mode. We confirm that the central value of the Kretschmann invariant against slow time at the origin also follows a universal curve with a period of $\Delta_I\simeq 0.09\pm 0.01$ for both of our $m=1$ families.

\begin{figure*}[!t]
    \centering    
    \vspace{0cm}
    \includegraphics[scale=0.19]{images/I_colorbar_m2_z.png} \,
    \includegraphics[scale=0.5]{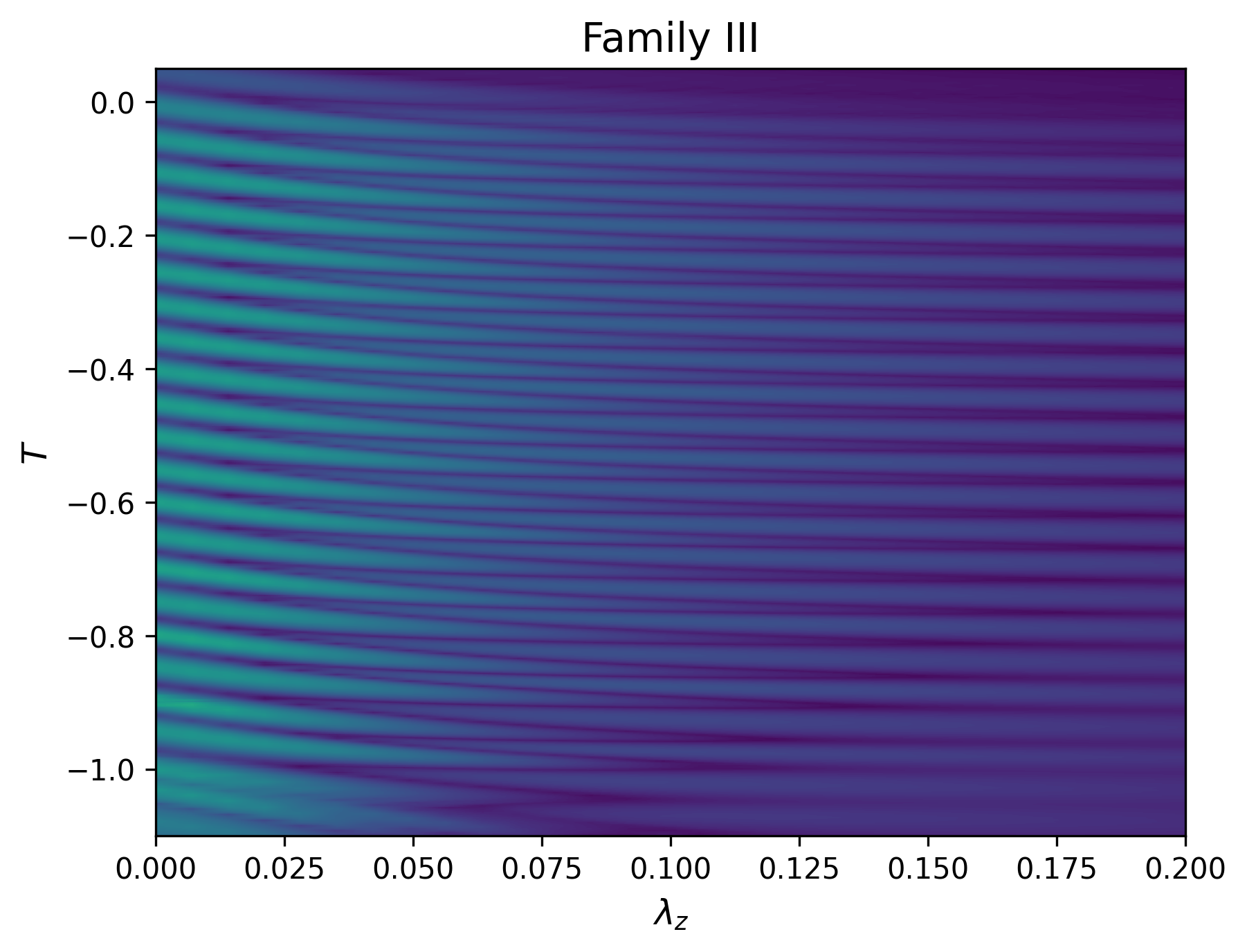} \,
    \includegraphics[scale=0.5]{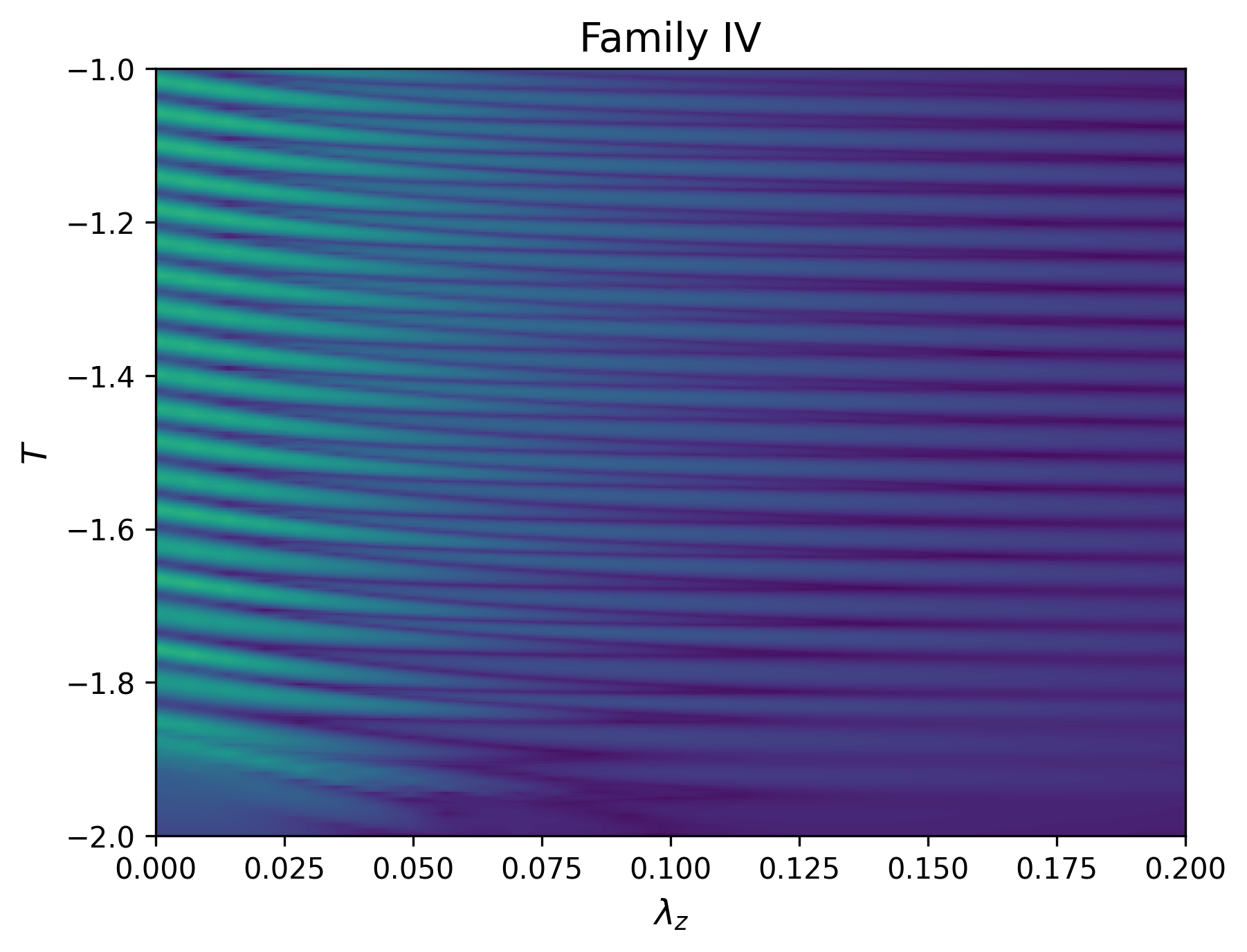} \,
    \caption{Color maps of the normalized Kretschmann invariant~$(\tau_*-\tau) \left|R_{abcd}R^{abcd}\right|^{\frac{1}{4}}$ for our best-tuned subcritical runs of Family III and IV initial data setups, all along the symmetry axis in single-null similarity coordinates.} 
\label{fig:m2_Kretsch_nullplots}
\end{figure*}

Similarly to Fig.~\ref{fig:m1_Kretsch_nullplots} for~$m=1$, Fig.~\ref{fig:m2_Kretsch_nullplots} displays the rescaled Kretschmann invariant for the best-tuned evolutions of families III and IV, using the same color scale for direct comparison. By construction the scalar field vanishes on the axis, but in this case fast enough that the Weyl tensor coincides with the Kretschmann invariant there. Therefore, the plotted quantity measures the trace-free part of the curvature on single null similarity adapted coordinates. In the~$m=2$ case, this contribution appears larger on the axis than for~$m=1$. Most importantly however, both panels of Fig.~\ref{fig:m2_Kretsch_nullplots} reveal a clear DSS structure over an extended spacetime region with~$\Delta \simeq 0.1$, and the fact that this behavior is observed across both families for each fixed mode supports the conclusion that universality holds in each case.

\begin{figure}[t!]
    \centering
    \includegraphics[width=\linewidth]{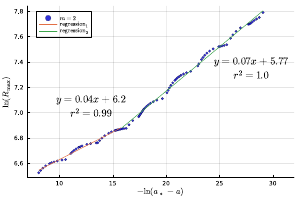}
    \caption{Global maximum of the Ricci scalar, $R_{\text{max}}$, as a function of the distance to criticality, $|a-a_*|$, for the $m=2$ family~III. 
The data follow a power law with two apparent slopes, but the one closer to criticality is taken as the relevant scaling. 
From this regime we extract $\gamma_{\textrm{sub}}\simeq 0.035$, a value compatible across all $m=2$ families (see Table~\ref{tbl:universalvalues}) and smaller than the $m=1$ result.}
    \label{fig:m2_sub_scaling}
\end{figure}

Looking into the phase space on the subcritical side, we analyze the scaling of $R_{\text{max}}$ (Fig.~\ref{fig:m2_sub_scaling}) of family III and obtain $\gamma_{\textrm{sub}}\simeq 0.035$. 
Interestingly, there are two distinct slopes in this figure, a behavior also present in the Einstein-Maxwell threshold case for~$l=2$, see Fig.~11 of~\cite{Reid:2023jmr}). If this were observed in yet more cases it may be good to build a model that captures the behavior.

A superposed oscillation (`wiggle'), a familiar signature of DSS critical solutions, is clearly visible in this figure. We determine the DSS period, extracted from this oscillation, by separating the data into the two regions with distinct slopes and fitting each segment with a sinusoidal function of the form~$f(x)=c+a\sin(\omega x + b) -2 \gamma x $. We find frequencies $\omega_1\simeq 2.19 \pm 0.16$ for the segment farer from the critical point and $\omega_2\simeq2.18\pm0.15$ for the near-critical one.
These frequencies agree within the error bars derived from the fit. The resulting DSS periods are being calculated using the relation $\omega = \frac{\Delta}{2\gamma}$ taken from Eq. (18) in~\cite{baumgarte2018aspherical}, since we expect that the maximum of the Ricci scalar scales according to the following relation
\begin{align}
    \ln\left(R_{\mathrm{max}}\right)= c-2\gamma\ln\left(a_*-a\right)+f\left(2\gamma \ln\left(a_*-a\right)+b\right).
\end{align}
In the first segment, this yields a DSS period
\begin{align}
    \Delta_1= 2\gamma_1\omega_1 \simeq 0.096 \pm 0.008,
\end{align}
while in the near-critical segment, we obtain a DSS period
\begin{align}
    \Delta_2= 2\gamma_2\omega_2 \simeq 0.15 \pm 0.01.
\end{align}
Surprisingly, the DSS segment of the segment farer from the threshold is in close agreement with the one measured by the central field values of the near-critical solution, $\Delta\simeq0.09$. The DSS period of the near-critical segment on the other hand is almost double that value, owing to the almost doubled value of the relevant slope. Additionally, observe that due to the sparse sampling in parameter space of the~$m=1$ case, Fig.~\ref{fig:m1_sub_scaling} of the present work does not reveal any periodic wiggle at all. Overall, we believe that the threshold picture remains consistent with that of the Choptuik solution where universality and DSS manifest in the spacetime.

\begin{figure}[t!]
    \centering
    \includegraphics[width=\linewidth]{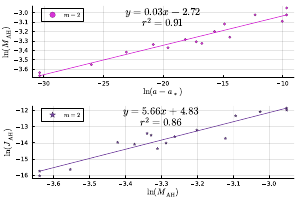}
    \caption{Supercritical scaling for the $m=2$ families. 
Top panel: First apparent horizon mass, $M_{\mathrm{AH}}$, versus the distance to criticality, $|a-a_*|$, for family~III. The data follow a power law with slope~$\gamma_{\textrm{sup}}\simeq 0.035$, in agreement with the subcritical exponent of Fig.~\ref{fig:m2_sub_scaling}. Bottom panel: Relation between the angular momentum and mass at the first apparent horizon, $J_{\mathrm{AH}}$ versus~$M_{\mathrm{AH}}$. While the data show scatter close to the threshold, there is a clear trend,$J_{\textrm{AH}}\propto M_{\textrm{AH}}^{5.66}$, which indicates that the spin parameter~$\chi_{\mathrm{AH}}=J_{\mathrm{AH}}/M_{\mathrm{AH}}^{2}$ decreases toward zero, in accordance with angular momentum being irrelevant and ruling out extremality in this family.\label{fig:m2_sup_scaling}}
\end{figure}

On the supercritical side, the horizon mass follows a power law with exponent $\gamma_{\textrm{sup}}\simeq 0.035$, averaging over both families, as demonstrated in Table.~\ref{tbl:universalvalues}. The top panel of Fig.~\ref{fig:m2_sup_scaling} shows the mass scaling for family~III, which is compatible with the subcritical exponent obtained near the threshold in Fig.~\ref{fig:m2_sub_scaling}. These values differ from those of the $m=1$ case, showing that the critical exponents themselves are mode-dependent. 

Taken together, these results demonstrate that while the specific values of $\Delta$ and $\gamma$ vary with $m$, each fixed-$m$ sector displays consistent behavior across different families, establishing universality within a given angular mode. 

The bottom panel of Fig.~\ref{fig:m2_sup_scaling} shows $J_{\mathrm{AH}}$ as a function of $M_{\mathrm{AH}}$ for the $m=2$ family~III.
As in the $m=1$ case, the data become scattered close to threshold, but the overall trend that the angular momentum~$J_{\mathrm{AH}}$ rapidly decreases with decreasing~$M_{\mathrm{AH}}$ is again clear. Approaching the threshold, for vanishing $M_{\mathrm{AH}}$, the dimensionless spin $\chi_{\mathrm{AH}}=J_{\mathrm{AH}}/M_{\mathrm{AH}}^{2}$ goes to zero. Compared with~$m=1$, the falloff of~$\chi_{\mathrm{AH}}$ is even steeper, reinforcing that angular momentum is irrelevant for this mode within the families of initial data we have considered.

\subsection{Consideration of extremality}
\label{subsec:extremality}

In light of recent results suggesting, and even proving in certain settings, that extremal (charged or spinning) black holes can arise as threshold solutions in various models~\cite{kehle2024extremal} it is natural to conjecture that our axisymmetric~$m=1$ system might likewise approach an extremal configuration at the black hole threshold starting from certain families of initial data. This perspective motivates using the dimensionless spin as a diagnostic tool: for an extremal Kerr black hole one expects~$J/M^{2}=1$, so if extremality were approached continuously our data should converge toward this bound on the supercritical side, revealing a limiting curve analogous to the extremal Reissner-Nordström case. In our present simulations, however, we find no evidence for a tendency towards extremal black holes, at least within the families that we investigated. More specifically, $\chi_{\mathrm{AH}}=J_{\mathrm{AH}}/M_{\mathrm{AH}}^{2}$ decreases rather than approaching unity near threshold. For this model, numerical evidence suggests universality at the threshold with vanishing angular momentum, and with a large basin of attraction. Therefore very different families of initial data ought to be considered to approach the extremal regime.

As an alternative means of assessing extremality at the black-hole threshold, we draw again on the perturbative framework of Gundlach and Baumgarte~\cite{PhysRevD.94.084012}. As discussed above, they developed a perturbative model of Type~II critical collapse with angular momentum, relating the mass and angular momentum scaling exponents to the eigenvalues of the critical solution’s linear perturbations.

In their framework the axial (rotational) mode with eigenvalue~$\lambda_{1}$ controls the scaling of~$J$ relative to that of~$M$, with $\lambda_{1}<0$ indicating that angular momentum is irrelevant and hence~$J/M^{2}\to 0$ at the threshold, which we observe empirically. Conversely, a marginal mode ($\lambda_{1}\approx 0$) would allow~$J/M^{2}$ to approach a non-vanishing finite value, signaling a potentially extremal limit, while~$\lambda_{1}>0$ would point to a relevant rotational mode and to rotating solutions at the threshold. This interpretation suggests that the sign of~$\lambda_{1}$ in our model likewise provides a natural diagnostic for extremality or non-extremality at the black-hole threshold. They find that only in the regime where~$\lambda_{1}/\lambda_{0}<0$, scaling relations can be derived for both~$J_{AH}$ and~$M_{AH}$ such that their scaling exponents can be related as 
\begin{align}
\label{eq:scaling_relation}
    \gamma_J =\tfrac{2-\lambda_1}{\lambda_0}=(2-\lambda_1)\gamma_{M}\,.
\end{align}
Where this scaling occurs, the dimensionless spin~$J/M^2$ itself scales as~$-\lambda_{1}/\lambda_{0}$ and therefore always vanishes as the threshold is approached. This can be understood from the fact that perturbations of the critical solution follow~$\lambda_0>0$ to either disperse or collapse, hence~$\lambda_{1}/\lambda_{0}<0$ implies~$\lambda_{1}<0$, which indicates a decaying axial perturbation and, looking at Eq.~\eqref{eq:scaling_relation}, implies that angular momentum decreases faster than the mass of the black holes. Our measured values, $\lambda_{0}\simeq9$ and~$\lambda_{1}\simeq-0.7$ for the~$m=1$ families and~$\lambda_{0}\simeq29$ and~$\lambda_{1}\simeq-4$ for the~$m=2$ families, lie in the regime of validity of the scaling relations and as expected belong to the case where the~$\lambda_1$'s are clearly negative, supporting the conclusion that angular momentum as described by the model is irrelevant.

Our working hypothesis is that, with our model and chosen initial data, extremality does not emerge at the threshold because this setup exhibits Type~II critical behavior with (empirically) a decaying axial perturbation with a fairly large co-dimension 1 basin of attraction. Kehle and Unger~\cite{kehle2024extremal} contrast their construction with the familiar Type~II critical phenomena discovered by Choptuik, in which the black-hole mass scales to zero and the critical solution is DSS. In their setting, by contrast, the critical solution is an extremal Reissner-Nordström black hole of finite mass and radius, with a single unstable mode.
Intuitively it may therefore be more easy to arrive numerically at the extremal segment of the threshold with models exhibiting Type~I critical collapse, where the black hole mass at the threshold is finite. Subsequently, one could attempt to increase the angular momentum that the data supports within a two-parameter family. The simplest adjustment to our setup to achieve this would be to use the established presence of Type I behavior in massive scalar-field models~\cite{Brady:1997fj, jimenez2022critical}. The field mass introduces a new length scale into the system that can effectively control the black hole mass at the threshold. If the amount of angular momentum could be controlled so as to persist in that regime, it would then perhaps be possible to approach an extremal limit. Although extremal critical collapse can belong to a completely different region of solution space, it might transition to Type~I or Type~II regions of the threshold. It would be very interesting to see then how the boson star solutions that usually exist at the threshold would change to make way for extremal solutions. Evidently much more work is needed in this direction to move away from mere speculation. 

\subsection{Discussion on gravitational wave content and competition of thresholds}

In our critical collapse study of massless complex scalar fields in twist-free axisymmetry~\cite{MaroudaCors2024}, we observed deviations from the spherical critical collapse picture, in agreement with what had been found with real scalar fields in axisymmetry~\cite{baumgarte2018aspherical, choptuik2003critical}. For sufficiently large asphericities we found a bifurcation of the centers of collapse and non-universal scaling exponents. Observing that these deviations shared features with vacuum critical collapse~\cite{baumgarte2023critical}, we conjectured that the deviations observed could be due to a competition between matter and vacuum black-hole formation thresholds, only possible with both matter and gravitational waves present, hence beyond spherical symmetry. Support for this was found by quantifying the matter and gravitational wave content in the strong-field region, in particular by looking at curvature scalars and positive-definite quantities associated with energy. Considering now the same matter model with a twisting Killing vector we can in principle study the threshold of rotating black holes in axisymmetry. In this setup however, we see no evidence for deviations from the standard picture of critical collapse, despite the asphericity of the data. We thus consider now the possible competition between thresholds in the twisting data. Our first step is again to quantify the gravitational wave content in the strong-field region. In order to assess the relative contribution of gravitational and matter degrees of freedom to the curvature, we also analyze the ratio between the maxima of the Weyl invariant and the Kretschmann scalar. The Kretschmann scalar is defined as
\begin{align}
    I = R_{abcd}R^{abcd}\,,
\end{align}
while the Weyl invariant is
\begin{align}
    W = C_{abcd}C^{abcd}\,,
\end{align}
with~$R_{abcd}$ the Riemann tensor and~$C_{abcd}$ the Weyl tensor. Because of the dynamical coupling between the two, this offers only an imperfect comparison between the gravitational wave and matter contributions to the collapse, but we are not aware of any better measure.

\begin{figure}[t!]
    \centering
    \includegraphics[width=\linewidth]{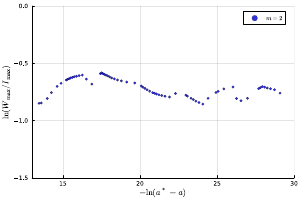}
    \caption{Ratio between the overall maxima of the Weyl invariant and the Kretschmann scalar, $W_{\mathrm{max}}/I_{\mathrm{max}}$, from data of family III, as a function of the distance to the threshold, $|a-a_*|$. The ratio remains approximately constant with some periodic features across the near-critical regime, with an average value~$\ln(W_{\mathrm{max}}/I_{\mathrm{max}}) \approx -0.71$. }
    \label{fig:maxW_vs_maxI}
\end{figure}

In Figure~\ref{fig:maxW_vs_maxI} we plot~$W_{\mathrm{max}}/I_{\mathrm{max}}$ as a function of the distance to the threshold. The ratio remains nearly constant throughout the near-critical regime, with an average value $W_{\mathrm{max}}/I_{\mathrm{max}}\approx 1/2$, indicating that the Weyl invariant accounts for a significant fraction of the total curvature, and hence that gravitational-wave content is a major component of the spacetime solution. In the figure, some structure is also observed, yet does not resemble a clear DSS pattern. We attribute this to the fact that the overall maxima in each evolution probably do not occur exactly at the same points for the Weyl and for the Kretschmann invariants. Ideally, the maximum of the pointwise ratio would provide a more accurate measure, and we have every reason to expect that clear DSS behavior would also appear there, as confirmed by the families studied (see~Figs.~\ref{fig:m1_Kretsch_nullplots} and~\ref{fig:m2_Kretsch_nullplots}). Nevertheless, the ratio of the overall maxima yields qualitatively similar conclusions regarding the gravitational-wave content.

The behavior observed in this class of axisymmetric solutions with substantial gravitational-wave content demonstrates that, despite aspherical configurations in which a more subtle story appears~\cite{choptuik2003critical,baumgarte2018aspherical,baumgarte2023critical}, the standard picture of critical collapse is emphatically not just an artifact of spherical symmetry. Importantly, the fact that the ratio~$W_{\mathrm{max}}/I_{\mathrm{max}}$ does not increase toward the threshold mirrors what is seen in the spherical Choptuik solution, and suggests that the gravitational wave content seems indifferent to the distance to the threshold (see Fig.~11 of~\cite{MaroudaCors2024} for comparison). Moreover, bifurcation that had been observed in the~$m=0$ case is not present here, perhaps due to angular momentum forcing everything to spiral around the symmetry axis and concentrate around the center. In line with our earlier hypothesis, the absence of bifurcation together with the constant ratio of Weyl to Kretschmann invariants shows that gravitational-wave content, though significant, does not dominate the threshold dynamics more in the limit of better tuning.

As we review in Appendix~\ref{app:no_competitor}, there is no vacuum threshold solution with angular momentum in axisymmetry with a single apparent horizon. Therefore, non-vacuum threshold solutions {\it with} angular momentum in axisymmetry are not close in solution space to a potential gravitational wave competitor. This observation may help explain why it is possible to describe accurately angular momentum at the threshold~\cite{PhysRevD.94.084012}, despite open challenges elsewhere in axisymmetry. In the symmetry class we studied here however, we saw that angular momentum on the horizon tends to zero as the threshold is approached. Not much is known about the threshold of collapse for twisting vacuum axisymmetric spacetimes, but they are expected to exist and therefore serve as a candidate competitor for the solutions we uncovered. Suppose that the twisting vacuum threshold were as complicated as in the non-twisting case (that is, if it presented deviations from the standard picture of critical collapse) and that its near-critical solutions were nearby in solution space to the solutions we computed here. Then, the universal self-similar behavior we observed here would, presumably, be suppressed. If not, the conjectured competition between the matter and gravitational wave dynamics would offer no explanation for the lack of deviations observed and would hence have to be augmented or directly abandoned. Study of the twisting vacuum threshold is therefore now a priority.

\section{Conclusions}
\label{sec:conclusions}

In this work we performed the first axisymmetric simulations of massless complex scalar field collapse with twist and higher ($m=2$) angular modes. We focused, as in earlier work, on initial data of pure~$l=|m|$ harmonics in the scalarfield. For this purpose we used the pseudospectral \textsc{bamps} code. The implementation of the~$m$-cartoon symmetry reduction together with a generalization to our apparent horizon finder enabled us to probe the near-critical regime very closely for families with~$m=1$ and~$m=2$. For~$m=1$ we recovered the DSS solution with~$\Delta\simeq 0.42$ and~$\gamma\simeq 0.11$, in agreement with~\cite{choptuik2004critical}. For~$m=2$ we identified for the first time a distinct DSS solution with smaller critical parameters, $\Delta\simeq 0.09$ and~$\gamma\simeq 0.035$, showing that universality holds within fixed-$m$ families but that the critical exponents depend on the angular mode. The phase space plot of the $m=2$ family exhibits two different slopes, but we only consider the one closer to the threshold for the calculation of the scaling exponent. In the same plot, we observe the presence of a periodic wiggle, with an estimated DSS period~$\Delta_1\simeq0.096\pm0.008$ for the segment farer away from the threshold of collapse and a DSS period $\Delta_2\simeq 0.15 \pm 0.01$ for the near-ciritical segment.

The relation between angular momentum and mass at horizon formation confirms that the dimensionless spin~$\chi_{\mathrm{AH}}$ decreases toward zero in both cases, demonstrating that angular momentum is irrelevant at the threshold with the setups that we have considered. This result is compatible with the general trend observed in~$m=1$ data of~\cite{choptuik2004critical}, but we obtain different scaling in this case. It is difficult to judge the exact cause of this, but possible explanations include the foliation dependence of quasilocal quantities or simply numerical error. Since we lack an alternative measure, for now we must make do. It would be desirable to make a systematic comparison of spacetime data computed independently for such spacetimes, but in view of our vacuum comparison~\cite{baumgarte2023critical} and the good agreement of our critical parameters with those of~\cite{choptuik2004critical}, we remain confident in the general picture painted by our data.

We interpreted our data with the language of the model of~\cite{PhysRevD.94.084012}, in which perturbations of an assumed self-similar threshold solution are described with two mode solutions and their respective Lyapunov exponents. In the context of the model, the negative subdominant Lyapunov exponent~$\lambda_1$ provides further evidence against extremality in this region of solution space. Considering curvature-invariants, we found that that the ratio~$W_{\max}/I_{\max}$ remains essentially constant near the threshold. In contrast to our earlier~$m=0$ evolutions~\cite{MaroudaCors2024} this suggests that the relative importance of  gravitational wave and matter contributions to the collapse is fixed. More speculatively, the value of the ratio itself indicates that the matter content is the primary driver.

In view of this, the critical solutions presented above are {\it very} interesting, even putting aside the fascinating question of how to choose initial data to arrive at configurations with substantial, even extremal, angular momentum at the threshold. In accordance with various studies in twist-free axisymmetry, in our earlier work~\cite{MaroudaCors2024} with the same matter model we found significant deviations from the standard picture of critical collapse as the degree of asphericity was increased. Yet here in the twisting case we unambiguously recover results, such as universal power-law scaling and DSS threshold solutions, familiar from the spherical setting. And we find no evidence for the formation of disjoint horizons or bifurcate centers of collapse. Why not? One may have thought that since threshold solutions here have non-vanishing angular momentum, the axisymmetric vacuum threshold, which cannot support angular momentum, could not serve as a nearby competitor in solution space. This would suppress one channel of interference suspected near the threshold in the non-twisting setting with general aspherical data. But in the limit of infinite tuning we arrived at solutions with vanishing angular momentum on the horizon, and we have no reason to believe that there is no twist-dominated vacuum axisymmetric threshold solution. This strongly motivates the investigation of vacuum solutions with twist.

\section*{Acknowledgments}

We are grateful to T.~Baumgarte, B.~Br\"ugmann, T.~Giannakopoulos, C.~Gundlach, D.~Nitzschke, R.~Pinto Santos and U.~Sperhake for helpful discussions and feedback on various aspects of the work. 

We acknowledge financial support provided by FCT/Portugal through grants 2022.01324.PTDC, UID/99/2025. D.\,C.\@ acknowledges support from the STFC Research Grant No.~ST/V005669/1, DiRAC projects ACTP284 and ACTP238, and from the Leverhulme Trust Early Career Fellowship ECF-2025-424.
H.\,R.\,R.\@ acknowledges financial support provided under the European Union’s H2020 ERC Advanced Grant ``Black holes: gravitational engines of discovery'' grant agreement no.\@ Gravitas-101052587.  Views and opinions expressed are, however, those of the authors only and do not necessarily reflect those of the European Union or of the European Research Council.  Neither the European Union nor the granting authority can be held responsible for them.  
A.\,V.\,V.\@ thanks FCT for funding with DOI 10.54499/DL57/2016/CP1384/CT0090, as well as support from the Universitat de les Illes Balears (UIB); the Spanish Agencia Estatal de Investigación grants PID2022-138626NB-I00, RED2022-134204-E, RED2022-134411-T, funded by MICIU/AEI/10.13039/501100011033 and the ERDF/EU; and the Comunitat Autònoma de les Illes Balears through the Conselleria d'Educació i Universitats with funds from the European Union - NextGenerationEU/PRTR-C17.I1 (SINCO2022/6719) and from the European Union - European Regional Development Fund (ERDF) (SINCO2022/18146). The figures in this article were produced with~\textsc{Plots.jl}~\cite{christ2022plots}, \textsc{Matplotlib}~\cite{Hunter:2007}, A. Khirnov's~\textsc{nr\_analysis\_axi} package~\cite{Khirnov:nranaaxi} and \textsc{ParaView}~\cite{ayachit2015paraview,ahrens200536}. Several calculations were performed using the~xAct package~\cite{xAct} for Mathematica. Numerical simulations were performed at the Leibniz Supercomputing Centre (LRZ), supported by the project~\textsc{pn36je}. The authors thankfully acknowledge the computer resources, technical expertise and assistance provided by CENTRA/IST. Computations were also performed at the cluster ``Baltasar-Sete-Sóis''.

\appendix
\section{Vacuum rotating black hole threshold in axisymmetry}
\label{app:no_competitor}

We here give the details behind the impossibility of an axisymmetric vacuum threshold solution with angular momentum with a single trapped surface. The Killing vector~$\xi^a$ of an axisymmetric spacetime (see Eq.~\eqref{eq:Killing}) can be used to compute the Komar angular momentum
\begin{align}
\label{eq:Komar_J}
J := \frac{1}{16\pi} \oint_{\partial S_{t}} \nabla^a \xi^b dS_{ab}\,,
\end{align}
where~$S_{t}$ is a surface within the 3-dimensional spacelike hypersurface~$\Sigma_t$ that has the topology of a 2-sphere. If~$\Sigma_t$ and~$S_{t}$ have metrics~$\gamma_{ij}$ and~$q_{ij}$ respectively, with metric determinants~$\gamma$ and~$q$,~$n^a$ is the timelike vector normal to~$\Sigma_t$ and~$s^i$ is a vector within~$\Sigma_t$, outward normal to~$S_{t}$, then the 2-surface element is~$dS_{ab}=2n_{[a}s_{b]}\sqrt{q}dx^2$ and the 3-volume element is~$dS_a = d\Sigma_a= - n_a \sqrt{\gamma}d^3x$. Using the divergence theorem~\cite{Gou07} for an antisymmetric tensor -- as is~$\nabla^a \xi^b$ by virtue of the Killing equation --, one finds that~$\oint_{\partial S_{t}} \nabla^a \xi^b dS_{ab} =2 \int_{S_{t}} \nabla_b \nabla^a \xi^b dS_{a} $, such that~\eqref{eq:Komar_J} becomes
\begin{align}
    J &= \frac{1}{8\pi}\int_{S_{t}} \nabla_b \nabla^a \xi^b dS_{a}= -\frac{1}{8\pi}\int_{S_{t}} R_{ab}\, \xi^b n^a \sqrt{\gamma}d^3x \label{eq:J_R}\\ 
      &= -\frac{1}{8\pi}\int_{\Sigma_{t}} T_{ab}  \xi^b n^a \sqrt{\gamma}d^3x\,, \label{eq:J_M}
\end{align}
where for~\eqref{eq:J_R} we have used that, since~$\nabla^a \xi^b$ is antisymmetric,~$g^{bd}\nabla_b \xi_d = 0$ such that~$\nabla_b \nabla^a \xi^b = g^{ac}g^{bd}(\nabla_b \nabla_c \xi_d - \nabla_c \nabla_b \xi_d) = g^{ac}R_{cbe}^b\xi^e = R^{a}{}_{e}\,\xi^e$;~\eqref{eq:J_M} comes from the trace-reversed Einstein field equations assuming that~$\xi^a$ belongs to~$\Sigma_t$ and hence~$g_{ab}\xi^b n^a=0$. The surface of integration can be generalized in~\eqref{eq:J_M} -- which corresponds indeed to~\eqref{eq:J_spacetime} -- from~$S_{t}$ to~$\Sigma_t$ because the integrand reduces to the matter content~$T_{ab}$ and is hence independent of the choice of~$S_{t}$ within~$\Sigma_t$. In this sense, the Komar angular momentum is conserved.

Now let~$S_{t}$ contain a black hole with apparent horizon~$\mathscr{H}_{t}$ and let~$V_{t}$ refer to the area contained between~$S_{t}$ and~$\mathscr{H}_{t}$ (see~\cite{Gou07} page 117 for a picture), with boundary~$\partial V_{t} = \partial S_{t} \cup \partial \mathscr{H}_{t}$, surface element~$dV_{ab}$ and volume element~$dV_{a}=d\Sigma_a$. The divergence theorem of~$\nabla^a \xi^b$ on this surface gives
\begin{align}
2 \int_{V_{t}} \nabla_b \nabla^a \xi^b dV_{a} &=  \oint_{\partial V_{t}} \nabla^a \xi^b dV_{ab} \nonumber\\
&= \oint_{\partial S_{t}} \nabla^a \xi^b dS_{ab} + \oint_{\partial \mathscr{H}_{t}} \nabla^a \xi^b dS_{ab}^{\mathscr{H}}\,, \nonumber
\end{align}
such that the Komar angular momentum evaluated on~$V_t$ gives
\begin{align}
    J &= \frac{1}{16\pi} \Big( 2 \int_{V_{t}} \nabla_b \nabla^a \xi^b dV_{a} - \oint_{\partial \mathscr{H}_{t}} \nabla^a \xi^b dV_{ab}^{\mathscr{H}}\Big)\nonumber\\
    &= \frac{1}{16\pi}  \Big( 2 \int_{\Sigma_{t}} T_{ab}  \xi^b n^a \sqrt{\gamma}d^3x - \oint_{\partial \mathscr{H}_{t}} \nabla^a \xi^b dS_{ab}^{\mathscr{H}}\Big) \nonumber\\
    &= J_{\text{M}} + J_{\text{BH}}\,.\label{eq:J_M+BH}
\end{align}
The Komar angular momentum can therefore only be sourced by matter and/or by rotating black holes. 

In vacuum~$J_{\text{M}}=0$,~$J=J_{\text{BH}}$. Considering~\textit{regular} initial data -- one that does not contain a black hole --~implies $J^{t=t_{0}}=J^{t=t_{0}}_{\text{BH}}=0$. Then, by the conservation of~$J$, any black hole formed from axisymmetric vacuum regular data, whether or not there is twist, has to be non-rotating to satisfy~$J^{t=t_{0}}=J^{t=t_{f}}=J^{t=t_{f}}_{\text{BH}}=0$. Therefore, there cannot be a vacuum threshold of collapse with angular momentum in axisymmetry. (It is possible to have two rotating black holes whose angular momentum combine to cancel each other, but if this occurred at the threshold of collapse a suitable perturbation to the data localized at one of the two horizons would  break the symmetry. We therefore disregard this more fine-tuned scenario.)

The fact that there is no rotating vacuum threshold in axisymmetry can also be understood in a more physical way from the statement `gravitational waves do not carry angular momentum in axisymmetry' (see Eq.~(2.13) in~\cite{RuiAlcNun07} and consider that~$\mathsterling_{\xi}h_{ab} = 0$ for Killing vector~$\xi^a$ and linearized metric~$h_{ab}$), and hence cannot collapse into rotating black holes.

In this study,~$J^{t=t_{0}}_{\text{BH}}=0$,~$J^{t=t_{0}}=J^{t=t_{0}}_{\text{M}}\neq 0$, such that if a black hole forms and absorbs all the matter, it has to rotate to satisfy~$J^{t=t_{0}}=J^{t=t_{f}}=J^{t=t_{f}}_{\text{BH}}\neq0$. The threshold of formation of such axisymmetric rotating black holes can only be driven by matter as it could not have formed through vacuum, and would hence not be a scenario where our conjectured competition could take place.

However, if at~$t=t_{f}$ not all the matter was inside the black hole, it was around it or had dispersed to~$\mathcal{J}^{+}$, then~$J^{t=t_{f}}$ would need not be equal to~$J^{t=t_{f}}_{\text{BH}}$, and hence~$J^{t=t_{f}}_{\text{BH}}$ would not need to be non-zero, i.\,e.\@ the black hole would not rotate. This case would correspond to the threshold of non-rotating black holes, which can be driven by both matter and vacuum and hence could be the result of the conjectured competition.

All of this only holds for axisymmetry with twist: a vacuum threshold of rotating black holes exists in full 3d. According to the conjectured competition, the 3d threshold would compete with a matter one in general.

To see the role played by the twist~$\omega_a=\epsilon_{abcd}\xi^b\nabla^c\xi^d$, consider the norm of the Killing vector~$\lambda^2=g_{ab}\xi^a\xi^b$ and the Geroch decomposition of the metric~$h_{ab} = g_{ab} - \lambda^{-2}\xi_{a}\xi_{b}$ (see~\cite{rinne2013axisymmetric}). Decomposing~$\nabla_a\xi_b$ in terms longitudinal to~$\xi_a$ and terms transverse to it -- so contracted with the metric~$h_{ab}$ -- gives~$\nabla_a\xi_b = \lambda^{-2} \big( \frac{1}{2}\epsilon_{abcd}\xi^c\omega^d + \xi_{[a}\nabla_{b]}\lambda^2\big)$. Placing this into~\eqref{eq:Komar_J} gives
\begin{align}
\label{eq:Twist_J}
    J &= \frac{1}{16\pi\lambda^{2}} \oint_{\partial S_{t}} \Big( \frac{1}{2}\epsilon_{abcd}\xi^c\omega^d + \xi_{[a}\nabla_{b]}\lambda^2 \Big) dS^{ab}\nonumber\\
    &=\frac{1}{32\pi\lambda^{2}} \oint_{\partial S_{t}} \epsilon_{abcd}\xi^c\omega^d dS^{ab}\,,
\end{align}
where we have assumed~$\xi^a$ belongs to~$\Sigma_t$,~$\xi^a n_a = 0$, and is tangent to~$S_t$,~$\xi^a s_a = 0$, such that~$\xi^a dS_{ab}=\xi^b dS_{ab} = 0$. It can be seen in~\eqref{eq:Twist_J} that in twist-free axisymmetry there is no angular momentum. All black holes formed in the twist-free setup of~\cite{MaroudaCors2024} were accordingly non-rotating. The threshold there could hence be the result of a competition between the twist-free matter and vacuum axisymmetric non-rotating thresholds. There can thus still be twist without angular momentum, if both terms in~\eqref{eq:J_M+BH} vanish, so in the absence of a rotating black hole or in vacuum.

For completeness, we recall that one can use the extrinsic curvature~$K_{ab}= - \gamma_{a}{}^{c}\nabla_c n^b$,~$\nabla^a \xi^b = \nabla^{[a} \xi^{b]} $ and~$\xi^a n_a = 0$ to obtain, for coordinates~$(t, r, \theta, \phi)$, a surface~$S_t$ with constant radius~$r$, and vector components~$s^i= (s^r,0,0)$ and~$\xi^i=(0,0,\xi^{\phi})$,
\begin{align}
\label{eq:appJAH}
  J &=\frac{1}{8\pi}\oint_{\partial S_{t}}K_{ij} \xi^j s^i \sqrt{q} d^2 x =\frac{1}{8\pi} \oint_{\partial S_{t}}K_{r\phi} \xi^{\phi} s^r \sqrt{q} d^2 x\,.
\end{align}
as in the calculation of the angular momentum for an apparent horizon~\eqref{eq:JAH}. Inserting the values of the Kerr metric, with~$a$ the Kerr parameter and~$m$ the mass of the Kerr black hole, and taking the limit~$r\rightarrow\infty $ (see~\cite{Gou07} page 123 for details) correctly gives~$J=J_{\text{Kerr BH}}=a\,m$, since for Kerr~$T_{ab}=0$. 

\bibliography{main.bbl}

\end{document}